\documentclass{article}
\usepackage{color}
\usepackage{setspace}
\usepackage[margin=1in]{geometry}
\usepackage{authblk}
\usepackage{float}
\usepackage{multirow}
\usepackage{multicol}
\usepackage{amsmath}
\usepackage[pdftex]{graphicx}
\usepackage{amssymb}
\def\beno{\begin{eqnarray*}}
\def\eeno{\end{eqnarray*}}
\def\be{\begin{eqnarray}}
\def\ee{\end{eqnarray}}
\renewcommand{\vec}[1]{\boldsymbol{#1}}
\usepackage{color, colortbl}
\usepackage[table]{xcolor}
\definecolor{Tgray}{rgb}{.9,.9,.9}
\usepackage{hhline}
\usepackage{algorithm2e}
\usepackage{natbib}
\begin{document}
\title{Genomic Region Detection via Spatial Convex Clustering}
\author{John Nagorski\thanks{jn13@rice.edu}}
\affil{Department of Statistics, Rice University}
\author{Genevera I. Allen \thanks{gallen@rice.edu}}
\affil{Department of Statistics, Rice University}
\affil{Department of Electrical Engineering, Rice University}
\affil{Department of Pediatrics-Neurology, Baylor College of Medicine}
\affil{Jan and Dan Duncan Neurological Research Institute, Texas Childrens Hospital}
\date{}
\maketitle
\begin{abstract}
{Several modern genomic technologies, such as
  DNA-Methylation arrays, measure spatially registered
  probes that number in the hundreds of thousands across multiple
  chromosomes.  The measured probes are by themselves less
  interesting scientifically; instead scientists seek to
  discover biologically interpretable genomic regions comprised of
  contiguous groups of probes  
which may act as biomarkers of disease or serve as a 
dimension-reducing pre-processing step for downstream analyses.  
  In this paper, we introduce an unsupervised feature learning technique
which maps technological units (probes) to biological units (genomic
regions) that are common across all subjects.   
We use ideas from fusion penalties and convex
clustering to introduce a method for Spatial
Convex Clustering, or SpaCC.  Our method is specifically tailored to
detecting multi-subject regions of methylation, but we also test our
approach on the well-studied problem of detecting segments of copy
number variation.   We formulate our
method as a convex optimization problem, develop a massively
parallelizable algorithm to find its solution, and introduce automated
approaches for handling missing values and determining tuning
parameters.  Through simulation studies based on real methylation and
copy number variation data, we show that SpaCC exhibits significant
performance gains relative to existing methods. Finally, we illustrate
SpaCC's advantages as a 
pre-processing technique that reduces large-scale genomics data into a
smaller number of 
genomic regions through several cancer epigenetics case
studies on subtype discovery, network estimation, and
epigenetic-wide association.}
\end{abstract}

\section{Introduction}
\label{sec1}
Modern genomic technologies take fine-grained measurements on
on human subjects that allow for increasingly individualized treatment
options for various diseases. 
Several of these technologies capture genomic information at spatially
registered locations on the DNA sequence; examples include point
mutations, next generation sequencing, copy number variation, and the
focus of this paper, 
DNA Methylation arrays which measure epigenetic variation.   
Here, the units returned by the technology, CpG sites, are not of
primary interest to scientists. 
More important are regions of CpG sites whose
cumulative impact affects gene function ~\citep{das2004dna}. 
To this end, we introduce an unsupervised feature learning technique
which maps technological units to biological units by coalescing
probes into contiguous genomic regions that are common across multiple
subjects.

Epigenetic technologies measure genetic aspects that affect gene regulation
beyond gene expression and transcriptomics.  
One such example is DNA Methylation, which measures the addition of a
methyl group to CpG sites creating  
5-methylcytosine.  High methylation levels 
(hypermethylation) have been shown to block gene transcription in cancer.
Similarly, low methylation levels (hypomethylation), typically at a global 
level, have also been observed by cancer researchers ~\citep{das2004dna}.
Recent advances in whole genome bisulfite sequencing technology
now yield methylation intensity measurements at hundreds of thousands
of CpG sites across the genome~\citep{bibikova2009genome}. 
The ratio of methylated intensity to total intensity, the so-called beta-value, is returned as a measure 
of the DNA methylation level at each site. 
Such technologies interrogate genomic regions (such as gene promoter
regions) by taking measurements of multiple CpG sites  
in close spatial proximity; beta-values in these regions are often
strongly correlated, indicating that they behave functionally as a
unit~\citep{eckhardt2006dna}. 
These observations imply that probes which are close in genomic distance and which display 
similar methylation levels are better treated as functional units, or genomic regions.
Previous work on region detection in the context of methylation has focused on differentially methylated region (DMR) 
discovery. Approaches in this area have utilized both smoothing techniques ~\citep{hansen2012bsmooth,jaffe2012bump} and the concept of linkage disequilibrium ~\citep{shoemaker2010allele}. 
The task of DMR discovery is, however, an inherently supervised one.
In this work, we focus on the unsupervised discovery of genomic regions for methylation data.
We develop a method for grouping probes into genomic regions that leads to more
interpretable scientific measurements, improves the performance of
downstream analyses, and can thus serve as a dimensionality reducing
pre-processing step for methylation data.

Another example of grouping spatial genomics data is the well-studied problem of copy number segmentation. 
Copy number variation (CNV) is one measure of structural variability in the
genome that quantifies large scale insertions and deletions of genetic information across the DNA sequence.
By utilizing array-CGH technology, structural differences relative to a reference sample are quantified as the log ratio of intensities at various loci across the genome.
Such differences have been linked to various cancers such as breast and lung ~\citep{shlien2009copy}.
Similar to DNA methylation, copy number measurements at a particular
loci are typically of lesser interest than regions of gain or loss, which signal large-scale amplification or deletions ~\citep{taylor2008functional}.
As such, the problem of Copy Number Segmentation has received much attention and numerous methods exist 
for this type of analysis ~\citep{olshen2004circular,venkatraman2007faster,wang2005method,nowak2011fused,bleakley2011group}.
Due to both its well-studied nature and similarity to region detection in the case of methylation data, we consider the task of copy number segmentation 
as a benchmark, and show that our method performs competitively to many state of the art methods.

\begin{figure}[H]
  \centerline{
  \includegraphics[width=.43\linewidth]{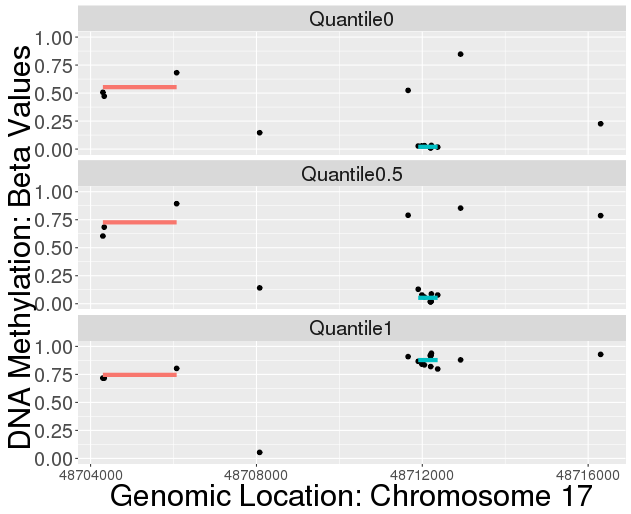} %
  \includegraphics[width=.55\linewidth]{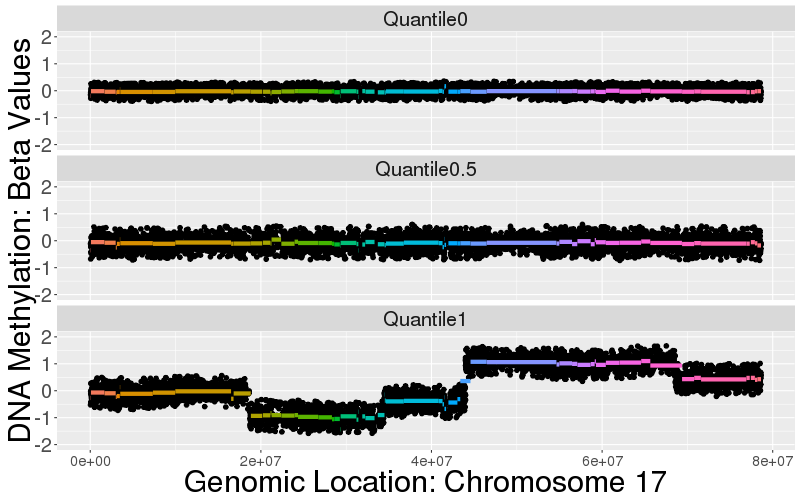}
  }
  \caption{SpaCC detected regions for TCGA Level 3 Breast Cancer
    DNA-Methylation Data $n=791, p=23515$ (left) and TCGA Level 2
    Ovarian Cancer Copy Number Variation (array-CGH) Data with
    $n=456,p=7658$ (right), both for Chromosome 17. For each data type
    we plot raw probe values overlayed with SpaCC clusters for three subjects.
    Subjects are ordered according to maximum deviation from the average cluster centroid over all subjects.
    For methylation data we note the ordering detects a hypomethylated region (blue), while
    for copy number data the ordering sorts patients according to genetic variability. In both cases, we note the discovery of common genomic regions across all patients.}
  \label{fig:ExampleRegions}
\end{figure}

For both methylation and CNV data, scientists seek entire
regions of CNV amplifications or deletions or regions of CpG sites
that behave as functional units.  For these regions to serve as
meaningful features for downstream multivariate statistical analyses,
they must be consistent across all subjects.  To address this, we
introduce an unsupervised learning 
method which maps technological units to more meaningful biological
units by clustering 
probes into genomic regions. 
An example of the genomic regions resulting from our method, Spatial
Convex Clustering (SpaCC), can be seen in Figure~\ref{fig:ExampleRegions}.
To detect such regions, we utilize the popular concept of fusion
penalties ~\citep{tibshirani2005sparsity} 
by extending methods for convex clustering to be appropriate for
spatial genomics data ~\citep{hocking2011clusterpath,ChiLan2014a}.  We
specifically focus on developing a fully automated and data-driven
method that can be used to pre-process spatial genomics data into genomic
regions for downstream analyses (Section~\ref{sec2}).  While we illustrate our method on
CNV data to benchmark our method against widely used segmentation
approaches (Section~\ref{cnvsim}), the main focus of this paper is on methylation data.  For
this, we show how our method can lead to improved results for
biomarker discovery in region-based epigenetic-wide association
studies (Section~\ref{subsection:apprewas}), and our resulting genomic regions can
yield improved features 
in downstream multivariate analysis such as clustering for subtype
discovery (Section~\ref{subsubsection:appmethsubtype}) and epigenetic network estimation (Section~\ref{subsubsection:appmethnetwork}).

\section{Spatial Convex Clustering}
\label{sec2}
Our objective is to discover groups of
probes which (i) have similar measurements, (ii) are spatially
contiguous, and (iii) are nearby on the chromosome. Taken together,
(i)-(iii) deliver  groups of probes similar enough to be considered as
single functional units, or genomic regions. Importantly, we require
that (i)-(iii) be satisfied simultaneously for all subjects, so that
the genomic regions discovered are intrinsically meaningful, and not
artifacts of a particular sample.  One approach to
multi-subject genomic region detection would be to consider clustering
the genomic probes based on the subject observations, as opposed to
the more commonly used clustering of observations.  To this end, we
merge ideas that use fusion penalties for copy number segmentation
~\citep{wang2005method,bleakley2011group,nowak2011fused} 
and the more recently introduced convex clustering methods
~\citep{ChiLan2014a,hocking2011clusterpath} to develop an automated
pipeline for  
feature learning with spatial genomics data. In particular, we address 
data-specific weighting schemes, automated methods for detecting the number 
and extent of clusters, as well as a principled approach for handling 
missing values.

\subsection{Optimization Problem and Algorithm}
Let $\vec X \in
\mathbb{R}^{n \times p}$ be our data matrix with $n$ subjects and $p$
variables (probes).
Our SpaCC problem is defined as 
\begin{equation} \label{eq:SpaCCProb}
\begin{aligned}
& \underset{\vec U \in \mathbb{R}^{n\times p}}{\text{minimize}}
& &  \frac{1}{2} \| \vec X - \vec U \|_F^2 + \gamma \sum_{i=1}^{p-1} w_{i}  \| \vec U_{\bullet i} - \vec U_{\bullet i+1} \|_2 \\
\end{aligned}
\end{equation}
Here, we use the Frobenius loss associated with the Gaussian
distribution to estimate the cluster means, given by the columns of
$\vec U \in \mathbb{R}^{n \times p}$.  The second regularization term
induces sparsity in the adjacent column differences of $\vec U$, thus forcing
adjacent column mean estimates to exactly fuse as the regularization parameter,
$\gamma$ is increased; fused columns of $\vec U$ form a
spatially contiguous cluster of variables.  As entire columns of $\vec
U$ are fused as a unit, the clusters are common to all samples.
The level of regularization is governed by
$\gamma$, with larger values encouraging more fusion and larger
spatial clusters, and the weights, $w_i$.  As discussed subsequently,
the latter are taken as spatial weights proportional to the inverse
distance between adjacent probes that are specific to each genomic
technology; these in turn, lead to more interpretable results and
computationally efficient algorithms.   Note that in contrast to
convex clustering methods ~\citep{ChiLan2014a,hocking2011clusterpath}
which cluster observations, our approach clusters variables;
additionally, we do not permit clusters among arbitrary  
groups of probes, and instead permit only genomically adjacent 
probes to coalesce together. In this sense, our method builds 
on the success of several existing fusion-based approaches that have
been proposed specifically for CNV
data~\citep{bleakley2011group,nowak2011fused}. 
Finally, note that we apply our SpaCC problem separately to data for
each chromosome.

To fit our SpaCC model, we adopt an approach introduced by
~\citep{ChiLan2014a} for 
convex clustering problems.  We reformulate \eqref{eq:SpaCCProb} by
introducing an auxiliary variable $\vec V$, where $\vec V_{\bullet,i}
= \vec U_{\bullet i} - \vec U_{\bullet i+1}$, and rewrite the penalty
in terms of  $\vec V$.  We then use the 
Alternating Minimization Algorithm (AMA)
\citep{tseng1991applications} optimization algorithm to fit our SpaCC
problem.  Formulating the augmented Lagrangian, 
we obtain updates for both primal variables ($\vec U_{\bullet i}$,
$\vec V_{\bullet i}$) and dual variables ($\vec \Lambda_{\bullet
  i}$), as shown in Algorithm~\ref{alg:Spacc}.  \\~\\ 
\begin{algorithm}[H]
 \KwData{$\vec X, \vec w, \gamma$}
 \KwResult{$\vec U, \vec V$ }
 \While{err $>$ tol}{
     $\vec U_{\bullet,i}^{k+1} = \begin{cases} 
      \vec X_{\bullet 1} + \vec \Lambda_{\bullet 1}^k & i = 1 \\
      \vec X_{\bullet i} + \vec \Lambda_{\bullet i}^k - \vec \Lambda_{\bullet i-1} & i \in \left\{2,\dots,p-1 \right\} \\
      \vec X_{\bullet p} - \vec \Lambda_{\bullet p-1}^k & i = p
\end{cases}$ \\
     $\vec V_{\bullet,i}^{k+1} =  \mathrm{prox}_{\gamma w_i / \nu\|.\|_2}(\vec U_{\bullet i}^{k+1} - \vec U_{\bullet i+1}^{k+1} - \frac{1}{\nu} \vec \Lambda_{\bullet i}^{k+1}) $\\
     $\vec \Lambda_{\bullet,i}^{k+1} = \vec \Lambda_{\bullet i}^k + \nu( \vec V_{\bullet i}^{k+1} - \vec U_{\bullet i}^{k+1} + \vec U_{\bullet i+1}^{k+1}) $\\
 }
 \caption{SpaCC AMA Algorithm} \label{alg:Spacc}
\end{algorithm}
~
\subsection{Spatial Weights}
An important input to our SpaCC problem is the weight vector, $\vec
w$, which incorporates genomic spatial structure and yields more
interpretable results and faster computation. Importantly, the choice
of weights is inversely proportional to the genomic distance between
adjacent probes.  While there are many such possible spatial weighting
schemes, we propose to use $w_i = \exp\left\{-\sigma d_i\right\}$,
where $d_i = \textrm{dist}(\textrm{probe}_i, \textrm{probe}_{i+1})$ is
the distance in basepairs between probes.
Here, $\sigma$ is chosen based on properties of genomic
data.   For example with copy number data, it is common to have
sizable portions of a chromosome amplified or
deleted~\citep{redon2006global}.  Hence, weights that
decay at a slower rate as a function of genomic distance will perform well
for CNV data; specifically, we take $\sigma = 0.00001$ in our
empirical studies.  For methylation data, we expect methylated regions
to form in small localized CpG islands near promoter regions of
genes~\citep{eckhardt2006dna}.  Hence, we take $\sigma = 0.0002$ for
methylation data, which yields a more rapid decay in weights as a
function of genomic distance.  Examples of these weight choices are
shown in the supplemental materials and the efficacy of these choices
are tested in simulations, Section~\ref{sims}, and real data
examples, Section~\ref{section:app}.  Overall, tailoring spatial
weights to specific technologies yields more interpretable genomic
regions, with larger clusters in CNV data and smaller localized
methylated regions, Figure~\ref{fig:ExampleRegions}.

Additionally, our choice of weights can dramatically decrease the
computational burden of our SpaCC 
algorithm.  Specifically, we apply hard-thresholding to the weights to
set tiny weights to zero.  Exact sparsity in the weight values
prevents distant adjacent probes from coalescing into a common genomic
region.  Further if $T$ weights are set to zero, the penalty term of
\eqref{eq:SpaCCProb} is perfectly separable into $T+1$ terms, yielding
$T+1$ subproblems.  These subproblems may in turn be solved in
parallel. For example using our methylation weight choice, the
Chromosome 17 TCGA Level 3 Breast Cancer Methylation SpaCC problem,
Figure~\ref{fig:ExampleRegions}, separates into $910$ subproblems that
can be solved completely in parallel.  This yields dramatic
computational gains when compared to 
the original problem which consists of $p= 23515$ probes.

\subsection{Missing Values and Parameter Selection}
We seek to develop a fully automated method for detecting
multi-subject genomic regions in CNV and methylation data; often
automated methods are more reliable and reproducible, as
practitioners have no knobs to tune that can yield different results
across studies and labs.  To
this end, we introduce two automated, optimization-based approaches to
handling practical problems with spatial genomics data: missing data
and regularization parameter selection.  First, missing values tend to
be a major problem for large-scale high-throughput genomics data.  For
example, the TCGA Breast Cancer Level 3
Methylation Chromosome 17 data shown in Figure~\ref{fig:ExampleRegions},
contains $14270$ total missing values. Similarly for TCGA Ovarian
Cancer Level 2 Copy 
Number Variation, Chromosome 17 contains $2349$ missing values.
Typically, missing genomics data is handled via a two-step process
where one first imputes the missing values using many popular off-the-shelf 
imputation routines \citep{troyanskaya2001missing} 
and then continues
with analyses of the fully imputed data.  This approach, however, is
less reproducible as results of downstream analyses can change
depending on the imputation procedure employed.  Instead, we propose
to fit our SpaCC model in the presence of missing data, effectively
eliminating the need to preform a separate imputation step.

As before, let $\vec X \in \mathbb{R}^{n\times p}$ be our data
matrix. Let $\mathcal{M} = \mathcal{M}_n \times \mathcal{M}_p \subset
\left\{1,\dots,n \right\} \times \left\{1,\dots, p \right\}$ denote
the indices of missing elements. We adopt an approach similar to that in
\citet{chi2016convex} 
to fit our SpaCC procedure in the presence of
missing values. Specifically, we fit our SpaCC loss function only over
the the non-missing elements of $\vec X$,
given by the indices $\mathcal{M}^C$:
\begin{equation}
\label{eq:SpaCCMiss}
\underset{\vec U \in \mathbb{R}^{n\times p}}{\text{minimize}}
\frac{1}{2} \sum_{j \in \mathcal{M}^C_p} \sum_{i \in \mathcal{M}_n^C}
( \vec X_{i j} - \vec U_{i j} )^2 + \gamma \sum_{i=1}^{p-1} w_{i}  \|
\vec U_{\bullet i} - \vec U_{\bullet i+1} \|_2 .
\end{equation}
First, notice that this optimization problem is still convex, and
hence our approach will yield the global solution.  We propose to
optimize \eqref{eq:SpaCCMiss} using the majorization minimization (MM)
algorithm \citep{hunter2004tutorial}. Defining the surrogate function
to be 
\begin{equation}
\label{eq:Major}
g(\vec U \mid \vec U^k) = \frac{1}{2} \left[ \sum_{j \in \mathcal{M}^C_p} \sum_{i \in \mathcal{M}_n^C} ( \vec X_{i j} - \vec U_{i j} )^2 +  \sum_{j \in \mathcal{M}_p} \sum_{i \in \mathcal{M}_n} ( \vec U_{i j} - \vec U^k_{i j} )^2 \right]+ \gamma \sum_{i=1}^{p-1} w_{i}  \| \vec U_{\bullet i} - \vec U_{\bullet i+1} \|_2,
\end{equation}
notice that $g()$ majorizes the objective $f(\vec U)$; namely (i)
$g(\vec U \mid \vec U^k) \geq f(\vec 
U)$ for all $\vec U$ and (ii) $g(\vec U^k \mid \vec U^k) = f(\vec
U^k)$. Iteratively
minimizing the surrogate objective creates a
non-increasing sequence of objective values $f(\vec U^k)$. Defining
$\vec T = \vec X_{\mathcal{M}^C} + \vec U_{\mathcal{M}}^k$, we augment
\eqref{eq:Major} leading to following algorithm to fit our SpaCC
problem in the presence of missing data:\\~\\ 
\begin{algorithm}[H]
 \KwData{$\vec X, \vec w, \gamma,\mathcal{M}$}
 \KwResult{$\vec U, \vec V$ }
 \While{err $>$ tol}{
    Set $\vec T = \vec X_{\mathcal{M}^C} + \vec U_{\mathcal{M}}^k$\\
    $\vec U^{k+1} \leftarrow \textrm{ Alg. 1  at } \vec T$
 }
 \caption{SpaCC Algorithm for Missing Data}
 \label{alg:missing}
\end{algorithm}
~
Notice that in contrast to traditional imputation routines,
Algorithm~\ref{alg:missing} 
does not explicitly replace missing elements in $\vec X$ prior to
analysis. Rather, missing indices are filled in iteratively via the
SpaCC cluster means for the genomic regions, $\vec U$.

Now that we have an automated way of fitting SpaCC with missing
values, we leverage this to propose a $k$-fold cross-validation scheme
to select the single tuning parameter, $\gamma$.  Note that $\gamma$
controls both the number of genomic regions and the extent of these
regions.  To select $\gamma$, we employ the approach of
\citet{wold1978cross} where we remove random elements of $\vec X$
in each fold and take the optimal $\gamma$ as the parameter whose
SpaCC imputed solution fit in the presence of missing values most
closely aligns with the removed data. Specifically, for
$k=1,\dots, \mathcal{K}$, where $\mathcal{K}$ the number of folds, we define the indices to be left out at the
$k$th fold to be $\mathcal{C}^k = \mathcal{C}^k_n \times
\mathcal{C}^k_p \subset \mathcal{M}^C_n \times \mathcal{M}^C_p$, where
$\bigcup_{k=1}^{\mathcal{K}} \mathcal{C}^k = \mathcal{M}^C$,
$\mathcal{C}^k \cap \mathcal{C}^{k \prime} = \emptyset$ and $\mid
\mathcal{C}^k \mid \approx \frac{1}{\mathcal{K}} \mid \mathcal{M}^C
\mid$, so that $\left\{ \mathcal{C}^k\right\}_{k=1}^{\mathcal{K}}$ is
an approximately equal sized partition of the non-missing elements of $\vec
X$. For each fold $k$, we introduce additional missing elements via
$\mathcal{C}^k$ and solve the missing data problem,
Algorithm~\ref{alg:missing}. Specifically, let $\mathcal{I}^k =
\mathcal{I}_n \times 
\mathcal{I}_p = (\mathcal{C}_n^k \cup \mathcal{M}_n^k) \times
(\mathcal{C}_p^k \cup \mathcal{M}_p^k)$. Given $\gamma$, we solve SpaCC
with missing values given by $\mathcal{I}^k$: 
\begin{equation}
\underset{\vec U \in \mathbb{R}^{n\times p}}{\text{minimize}} \frac{1}{2} \sum_{j \in (\mathcal{I}^k_p)^C} \sum_{i \in (\mathcal{I}_n^k)^C} ( \vec X_{i j} - \vec U_{i j} )^2 + \gamma \sum_{i=1}^{p-1} w_{i}  \| \vec U_{\bullet i} - \vec U_{\bullet i+1} \|_2
\end{equation}
We can solve this problem via Algorithm~\ref{alg:missing}, replacing
$\mathcal{M}$ with $\mathcal{I}^k$; we denote the solution as $ (\vec
U_{\gamma}^*)^k$. Then, we evaluate the performance of $\gamma$ on the $k$th
fold by comparing the solution $(\vec U_{\gamma}^*)^k$ to the left out
(non-missing) elements of $\vec X$ via mean squared error: $MSE(\gamma, k) = \sum_{i \in \mathcal{C}_n^k} \sum_{j \in \mathcal{C}_p^k} \left[ (\vec U_{\gamma}^*)^k_{ij}  - \vec X_{ij} \right]^2$
Our complete cross validation algorithm is given as follows:\\~\\
\begin{algorithm}[H]
 \KwData{$\vec X, \vec w, \gamma, \left\{ \mathcal{I}^k \right\}_{k=1}^{\mathcal{K}}$}
 \KwResult{$\vec U, \vec V$ }
 \For{$k = 1,\dots \mathcal{K}$}{
    \For{$\gamma = \gamma_1,\dots,\gamma_T$} {
        Algorithm~\ref{alg:missing} at $\lambda$ with missing indices given by $\mathcal{I}^k$ \\
        Compute MSE over $\mathcal{C}^k$
    }
 }
 \caption{SpaCC Algorithm for Cross Validation}
\end{algorithm}
~

\begin{figure}[H]
  \centerline{
  \includegraphics[width=.55\linewidth]{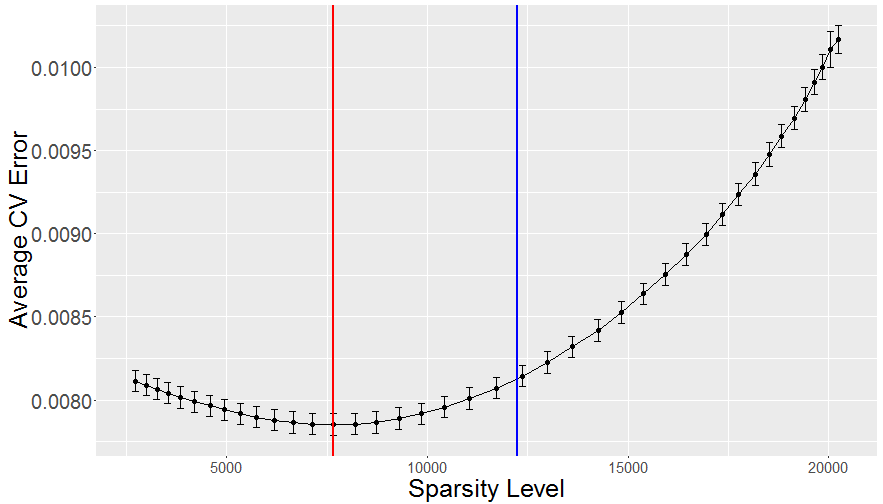} %
  \includegraphics[width=.52\linewidth]{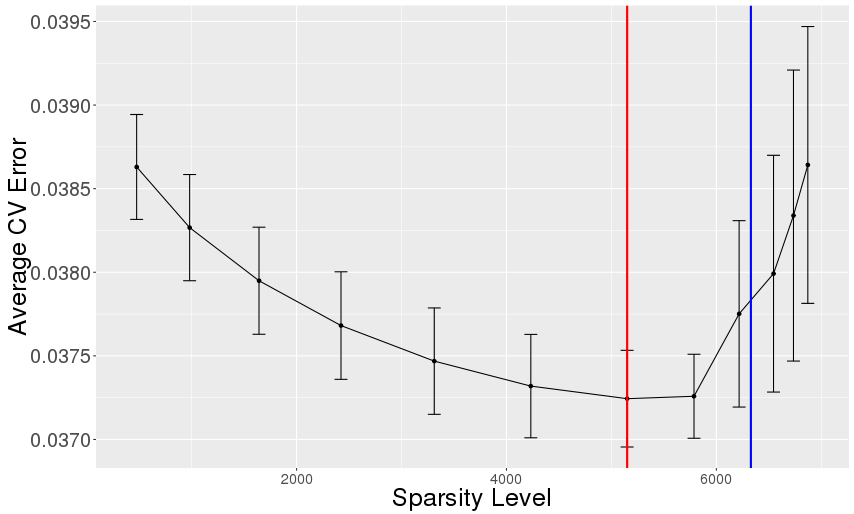}
  }
  \caption{Cross Validation plots for both methylation (left) and Copy Number (right). Sparsity-level is plotted along the x-axis and Average Error on the y-axis. The red line shows the sparsity-level of minimum Average Error, and the blue line shows sparsity level obtained after thresholding. }
  \label{fig:CV}
\end{figure}

Given the tuning parameter, $\gamma^*$, selected by minimizing the 
cross-validation error, we noticed that SpaCC identifies genomic regions 
which are spatially smaller than desired. Stated another way,
cross-validation tends to underestimate the sparsity level in the
differences,  
$\vec V$.  Such behavior is well known for the lasso and other sparse
problems and is 
hence why many advocate using the one-standard-error cross-validation
rule~\citep{van2011adaptive}. An alternative approach is to
post-process the results after cross-validation by thresholding
\citep{meinshausen2009lasso}. We adopt such a scheme and propose to
threshold the elements of $\vec V_{\gamma^*}$ at a level
proportional to the estimated noise, $\hat \sigma^2$; specifically we
threshold at the level, $| \left(\vec
V_{\gamma^*}\right)_{ij}| < \sqrt{\frac{\log(p)}{n}} \hat \sigma$,
similar to that proposed by
~\citet{meinshausen2009lasso}.
In Figure~\ref{fig:CV}, we give examples of cross-validation error curves over
a sequence of $\gamma$ values for the TCGA Ovarian Copy Number
and Methylation chromosome 17 example from
Figure~\ref{fig:ExampleRegions}; both the minimum cross-validation
error as well as our proposed thresholding level are shown.  The
efficacy of our cross-validation and post-thresholding procedure are
further studied in subsequent simulations.  Overall, we have developed
an automated and principled optimization-based approach for
handling missing values and selecting the number and extent of genomic
regions that offers many advantages over competing approaches.

\section{Simulation Studies}
\label{sims}

We study the performance of SpaCC empirically and compare our approach
to existing methods through simulations based on real array-CGH and
DNA-methylation data.  Before presenting our simulation results, it
will be helpful to introduce some notation.  Our probes are indexed by $j \in
\left\{1,2,\dots,p\right\}$.
Associated with each probe is a genomic location $l_j \in \mathbb{R}$, $j = 1,\dots,p$,
and we denote the distance between genomic locations by $d_{ij}$.
The probes are partitioned into clusters, $c_g$, indexed by $g=1,\dots, G$.
For each probe $j$ we denote the cluster to which it belongs via $r(j) = \left\{g \mid j \in
c_g \right\}$.
Our method and competitors will estimate a clustering
$\left\{ \hat c_g\right\}_{g=1}^{\hat G}$, with $\hat c_g \subset
\left\{ 1,\dots,p\right\}$ and $\hat c_{g} \cap \hat c_{g^{\prime}} =
\emptyset$. We will then evaluate the performance of all methods by
by comparing the estimated clustering
$\left\{ \hat c_g\right\}_{g=1}^{\hat G}$ to the true clustering
$\left\{ c_g\right\}_{g=1}^{ G}$ using common clustering metrics such
as the Rand, Adjusted Rand, and Jaccard Indexes
\citep{wagner2007comparing}. Note that
these clustering metrics are general and do not specifically account
for the fact that we require clusters to be spatially contiguous.
Hence, we also employ a lesser-known entropy-based clustering metric,
the Variation of Information (VI) metric \citep{meilua2007comparing}, that
is better suited to measuring the information loss between two sets of
clusters that may be nested, as we would often expect with spatially
contiguous clusters.

\subsection{Simulation Studies: Copy Number Segmentation}
\label{cnvsim}
We first evaluate the performance of SpaCC for the well-studied problem 
of copy number segmentation for array-CGH data.  While not our primary focus, 
the problem shares many attributes in common with the problem of genomic region 
detection for methylation data. These common features along with many 
popular tools and software packages ~\citep{dnacopy,cntools} 
make the problem an ideal benchmark to test the performance of our
method.

Our simulations are based on TCGA
Ovarian Cancer Level II array-CGH data for chromosome 17 ~\citep{cancer2011integrated}, where we adopt the
probe locations and use data for observed subjects to form the mean
simulated CNV signal as well as the locations of copy number
segments with amplifications or deletions. Specifically, we simulate
series for $j=1 \ldots p$ probes and $i =1 \ldots n$ subjects from the
following model: $X_{ij} = \mu_j + s_{i,r(j)} m_{i,r(j)} + \epsilon_{ij}$.
Here, $\mu_j$ is the base mean of the series which is taken from a
subject in the TCGA data that was detected as having no amplifications
or deletions according to DNACopy \citep{dnacopy}; $s_{i,r(j)}$ is an
indicator of an amplification or deletion for subject $i$ in region
$r(j)$ where the regions $r(j)$ are taken from the observed segments
as detected by DNACopy for subjects from the TCGA data; $m_{i,r(j)}$
gives the mean shift for the amplification or deletion in region
$r(j)$; and $\epsilon_{ij} \sim N(0, \sigma)$ is iid additive noise.  As not all
subjects will have copy number changes for each region, we simulate
$s_{i,r(j)} \sim Bernoulli(q)$.  Also, as each subject could have an
amplification or deletion of differing magnitude, we simulate
$m_{i,r(j)} \sim \pm U(a,b)$.

We study four simulation scenarios of varying difficulty.  First, we
study the effect of the magnitude of the copy number changes relative
to the noise level by taking $a=0.2,b=0.4,\sigma=.1$ for large changes,
and $a=0.05,b=0.3,\sigma=.1$ for small magnitude shifts that are more
difficult to detect.
Next, we use two different subsets of DNAcopy regions detected for a real TCGA patient, with
easier or more difficult to detect shapes to seed the segment boundaries;
that is, easier to detect segments tend to be spatially
longer and harder to detect segments are shorter and more fragmented.
Note that the easy and hard set of segments are shown in the
Supplemental Materials. Also related to shape, we vary the probability of a shift, $q=0.7,0.5$,
with a larger probability corresponding greater consensus across subjects, and hence easier shift detection.
\begin{table}[H]
{\small
\hspace*{-.5in}\begin{tabular}{| c | c | c | c | c | c | c | c | c | c |}
\hline
& \multicolumn{3}{c|}{Rand} & \multicolumn{3}{c|}{Jaccard} & \multicolumn{3}{c|}{Variation of Information} \\
\hhline{~---------} & SpaCC & DNACopy & CNTools & SpaCC & DNACopy & CNTools & SpaCC & DNACopy & CNTools \\
\hline
\hline
Easy(S)-Easy(M) & \cellcolor{Tgray} .98 {\small(.03)} & .87 {\small(.02)} & .94 {\small(.02)} &\cellcolor{Tgray} .94 {\small(.09)} & .76 {\small(.03)} & .77 {\small(.12)} &\cellcolor{Tgray} .06 {\small(.11)} & .38 {\small(.05)} & .43 {\small(.18)}\\
Easy(S)-Hard(M) & \cellcolor{Tgray} .99 {\small(.02)} & .88 {\small(.02)} & .93 {\small(.02)} & \cellcolor{Tgray}.97 {\small(.06)} & .76 {\small(.03)} & .71 {\small(.11)} & \cellcolor{Tgray}.02 {\small(.07)} & .39 {\small(.05)} & .63 {\small(.19)} \\
Hard(S)-Easy(M) &\cellcolor{Tgray} .99 {\small(.00)} & .89 {\small(.01)} & .85 {\small(.02)} &\cellcolor{Tgray} .99 {\small(.01)} & .73 {\small(.03)} & .38 {\small(.09)} &\cellcolor{Tgray} .00 {\small(.02)} & .39 {\small(.04)} & 1 {\small(.18)} \\
Hard(S)-Hard(M) & \cellcolor{Tgray}.99 {\small(.00)} & .89 {\small(.01)} & .85 {\small(.01)} & \cellcolor{Tgray} .99 {\small(.02)} & .72 {\small(.03)} & .37 {\small(.07)} &\cellcolor{Tgray} .00 {\small(.03)} & .43 {\small(.04)} & 1.15 {\small(.17)} \\
\hline
\end{tabular}
\caption{CNV Segmentation performance of SpaCC, DNACopy, and CNTools
  over simulation regimes with easy or hard segment shapes (S) and easy or hard
magnitude shifts (M). }
\label{tab:cnvsim}
}
\end{table}
Results indicate that SpaCC outperforms according to all metrics on
all regimes in this simulation.  As illustrated subsequently for
real data, DNACopy segments each subject separately.  As
such, DNACopy performs well on per-subject basis, but performs poorly
at the detecting segments common across all patients.  To address this
shortcoming, the CNTools package and its implementation of the
Circular Binary Segmentation (CBS) algorithm offers an alternative
whereby common segments are returned for all subjects. Yet, CNTools'
still fails to reach
a reasonable consensus regarding common subject segments.  Given the
discrepancy between subject's amplification/deletion regions, CNTools
tends to deliver shorter segments, wherein agreement can be reached
across several subjects. These short segments often partition the
true, larger segments, but are both difficult to interpret and
necessarily score poorly on the various metrics.  In contrast, SpaCC's
regularized solution yields an improved consensus between the
individual series and their common segments, detecting larger segments
more closely aligned with the underlying truth.

\subsection{Simulation Studies: Methylation Region Detection}
\label{ref:methsim}
Next we evaluate SpaCC's performance for the task of methylation region discovery.
We again model our study based on real data, utilizing TCGA Breast
Cancer Level 3 Methylation data for chromosome 17 ~\citep{cancer2012comprehensive}.
Initial clusters, $\left\{c_g \right\}$, are detected using SpaCC, which then act as the ground truth for what follows.
We simulate methylation beta values via the cdf transform $X_{ij} = \Phi(z_{ij})$, for subjects $i=1,\dots,n$ and probes $j=1,\dots,p$,
to ensure values in $(0,1)$. Our simulated methylation regions are
denoted by the degree of correlation of probes within the region.
Specifically, we take $z_{ij}$ to be $\vec z_i \sim N_p(\vec 0, \vec \Sigma)$.
with spatial covariance given by
$\Sigma_{ij} =
\begin{cases}
\exp\left\{ -d_{ij}/\sigma_w \right\} & r(i) = r(j)\\
\exp\left\{ -d_{ij} / \sigma_b \right\} & o.w.
\end{cases}$
where $d_{ij}$ is the distance between probes.  Here $\sigma_w$
controls the decay rate of the within cluster spatial correlation, and
$\sigma_b$ similarly between clusters. The difficulty of the
simulation is controlled by difference between the the within and
between spatial correlation decay rate. 
By varying the ratio of decay rates, $\sigma_b / \sigma_w$, we
consider three scenarios of high, medium, and low within region
spatial correlation, relative to between; these scenarios correspond
to $(\sigma_w,\sigma_b) = (100KB,.01KB), (100KB,.1KB), (100KB,1KB)$,
respectively. We compare SpaCC's performance to the linkage
disequilibrium method introduced in ~\citep{shoemaker2010allele}. 

\begin{center}
\begin{table}[H]
\hspace*{.15in}\begin{tabular}{| l | l | l | l | l | l | l | l | l |}
\hline
& \multicolumn{2}{c|}{Variation of Information}&\multicolumn{2}{c|}{Jaccard}&\multicolumn{2}{c|}{Rand}  &\multicolumn{2}{c|}{AdjRand}   \\
\hhline{~--------} &SpaCC & LD  &SpaCC & LD &SpaCC & LD& SpaCC & LD\\
\hline
High & \cellcolor{Tgray} .03 {\small(.00)} & .45 {\small(.00)} & \cellcolor{Tgray} .91 {\small(.00)} & .46 {\small(.00)} & .99 {\small(.00)} & .99 {\small(.00)} & \cellcolor{Tgray}.95 {\small(.00)} & .63 {\small(.00)} \\
Medium & \cellcolor{Tgray} .17  {\small(.00)} & .55  {\small(.00)} & \cellcolor{Tgray} .81 {\small(.00)} & .43 {\small(.00)} & .99 {\small (.00)} & .99 {\small(.00)} & \cellcolor{Tgray}.90 {\small(.00)} & .60 {\small(.00)}\\
Low & \cellcolor{Tgray} .52 {\small(.02)} & .76 {\small(.00)} & \cellcolor{Tgray} .60 {\small(.04)} & .37 {\small(.00)} & .99 {\small(.00)} & .99 {\small(.00)} & \cellcolor{Tgray}.75 {\small(.03)} & .54 {\small(.00)} \\
\hline
\end{tabular}
\caption{SpaCC and Linkage Disequilibrium performance for region
  detection on simulated methylation data. We see SpaCC outperforms
  linkage disequilibrium across all metrics and for all levels of
  difficulty, given by the degree of spatial correlation.} 
\label{tab:irc}
\end{table}
\end{center}
We again note that SpaCC outperforms existing region-detection methods
across all regimes and metrics. A portion of SpaCC's success may be
attributed to the continuous nature of its spatial fusions, wherein
clusters are joined in a smooth fashion via continuous spatial weight
decay. The LD method, by contrast, implements a greedy strategy for
accumulating clusters based on a discrete window size, here 500 bp.
This fixed window size gives the LD method less flexibility to detect
long range clusters when they are present; SpaCC, by not choosing
apriori distance cutoffs, does not have this difficulty.

\section{Applications of SpaCC to Cancer Genomics Data}
\label{section:app}

We now illustrate how SpaCC can be used as a pre-processing tool to
improve the analysis of real array-CGH and DNA-methylation data.
Using case studies from TCGA ovarian, breast, and lung cancers we
show how SpaCC can yield improved copy number segmentation results in
Section~\ref{subsection:appcnv}, but mainly focus on the more novel
application of 
detecting methylation regions.  For this, we show how SpaCC can yield
improvements in subtype discovery, Section~\ref{subsubsection:appmethsubtype}, inferring
epigenetic networks, Section~\ref{subsubsection:appmethnetwork}, and biomarker discovery in
region-based epigenetic-wide association studies (rEWAS),
Section~\ref{subsection:apprewas}.  As EWAS and especially rEWAS are
relatively new types 
of epigenetic analyses, we include a more careful study of this
application with additional simulation results in
Section~\ref{subsubsection:apprewassim} and an application to TCGA lung cancer data in Section~\ref{subsubsection:apprewaslung}.

\subsection{Ovarian Cancer Copy Number Segmentation}
\label{subsection:appcnv}

\begin{figure}[H]
  \centerline{
  \includegraphics[width=.4\linewidth]{SpaCC_CNVQuant_800x500.png}
  }
  \centerline{
  \includegraphics[width=.4\linewidth]{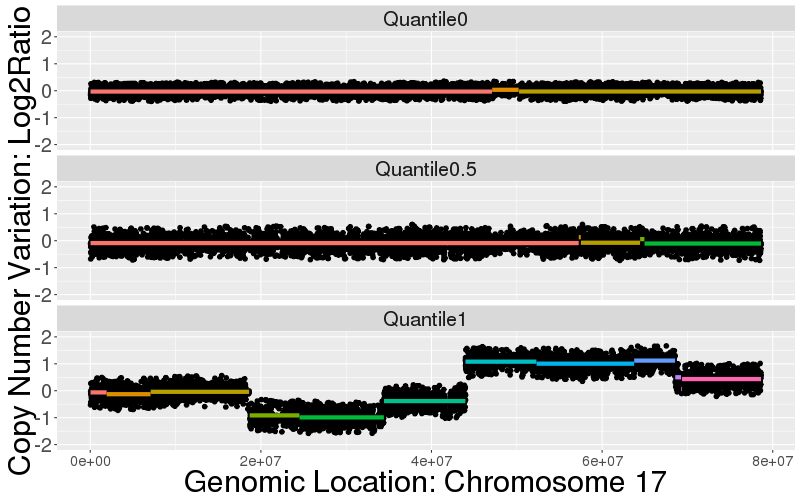}%
  \includegraphics[width=.4\linewidth]{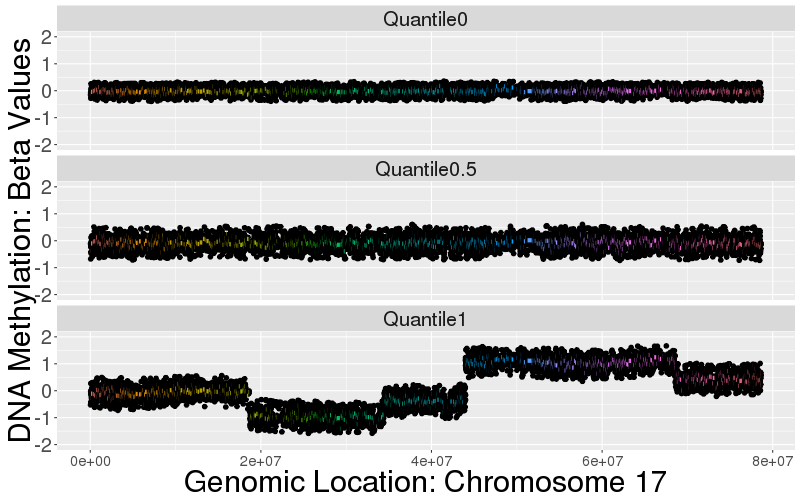}
  }
  \caption{Copy Number Segmentation results for TCGA Ovarian cancer
    chromosome 17.  We compare SpaCC (top) to DNAcopy
    (bottom left) and CNTools (bottom right). Copy number series over
    the chromosome are represented by black points and segments are
    represented by horizontal colored lines. The subjects with the
    minimum, median, and maximum copy number changes over the cohort
    are visualized.}
\label{figure:cnvappfig}
\end{figure}

We investigate SpaCC's segmentation performance on
Ovarian Cancer TCGA Level 2 Copy Number data ~\citep{cancer2011integrated}.
We report results for Chromosome 17, which is where the important BRCA1 gene,
which has been widely associated with ovarian cancer
~\citep{king2003breast}, 
is located.  The number of
probes totals $p = 7103 $ and we consider $n=456 $ subjects.  In
Figure~\ref{figure:cnvappfig}, we visually compare segments discovered by SpaCC to
those of the two most popular competitors, DNACopy \citep{dnacopy} and CNTools
\citep{cntools}. We plot three subjects' copy number series for the whole
chromosome and overlay the estimated segments, colored consecutively,
detected by each method; the level of each segment corresponds to the
subject's average copy number per segment.  The three subjects plotted
represent the subject with the minimum, median, and maximum copy
number changes across the cohort.

These results visually illustrate the many advantages of SpaCC over
existing tools for copy number segmentation. In particular, DNACopy, while
performing well on a per-subject basis delivers inconsistent segments
across multiple subjects. Indeed, the number of segments returned by
DNACopy varies from $1$ to $57$, depending on subject. Thus, the
segments produced by DNACopy cannot be used as a pre-processing step
before further multivariate analyses, as the
features are unaligned across the subjects. Conversely,
CNTools, while delivering segments consistent across the subjects, finds
difficulty assessing the trade-off between subject-specific patterns
and patterns common to all subjects. The result is an awkward
consensus with many ($3754$), short segments. This
"shattered" appearance makes interpretation difficult due to both the
number of segments and their size. In contrast, SpaCC finds longer
more interpretable segments ($61$), finding a more appropriate balance
between subject-specific and common patterns across the cohort. As
such, SpaCC is ideally suited as a pre-processing method to segment
and reduce the dimension of copy number data before further
multivariate analyses.

\subsection{Detecting Methylation Regions with SpaCC}
\label{subsection:appmeth}

\begin{figure}[H]
  \centerline{
  \includegraphics[width=.5\linewidth]{Methylation_Chr17RegionFigure.png} %
  \includegraphics[width=.5\linewidth]{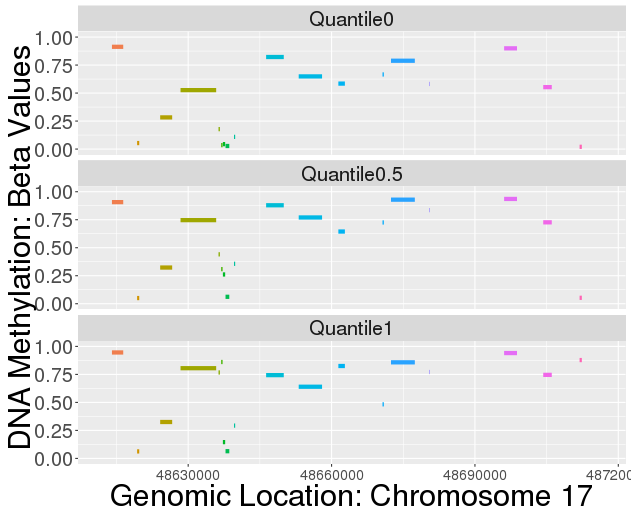}
  }
  \caption{Breast Cancer methylation region detection results.  A
    portion of Chromosome 17 with SpaCC-detected regions
    overlayed for the subjects with the minimum, median, and maximum
    variation in methylation levels (left).  We zoom in on the 48.70Mb
    - 48.72Mb region of
    chromosome 17 (right) containing the promoter region of the ABCC3
    gene.}
\label{figure:methappfig}
\end{figure}

We now study and illustrate how SpaCC can be used for biomarker
discovery and as a pre-processing tool that yields a reduced and more
meaningful set of features for downstream 
analysis of methylation data.  First, we apply SpaCC to discover
methylation regions from the TCGA Level 3 Breast Cancer data
~\citep{cancer2012comprehensive} 
for chromosome 17 with $n=791$ subjects and $p=23515$ probes.
We visually examine methylation regions returned by SpaCC in Figure
~\ref{figure:methappfig}.  Notice that the methylation
regions detected are
much shorter than their Copy Number counterparts. Segments in the
latter tend to be amplified or deleted in large chromosomal regions, whereas
regions of consistent methylation levels to occur in small localized areas
corresponding to promoter regions of genes ~\citep{eckhardt2006dna}.
On the right in Figure~\ref{figure:methappfig}, we can
easily see how SpaCC features aggregate across probes whose levels are
meaningful and consistent across all
subjects; this is illustrated by the region denoted in blue.
In contrast, SpaCC has not fused the two probes flanking this region
as their levels differ significantly from the blue region for the
subjects with the minimum and median variation in methylation levels shown.
These local regions allow us to capture fine-grained
characteristics that are more biologically meaningful than examining
single probes. For example, the highlighted blue region is
hypomethylated in most subjects, but
hypermethylated for the bottom subject shown, which has the maximum variation in
methylation levels.  This region corresponds to
the promoter region of the ABCC3 gene which is associated with HER2
and luminal breast
tumors ~\cite{o2008functional}.
In total, SpaCC detected $9,080$ methylation regions for chromosome 17,
which offers a reduction from the original $23,515$ probes.

\subsubsection{Breast Cancer Methylation Subtype Discovery}
\label{subsubsection:appmethsubtype}

\begin{figure}[H]
\centering
\begin{tabular}{ccc}
\includegraphics[width=.3\linewidth]{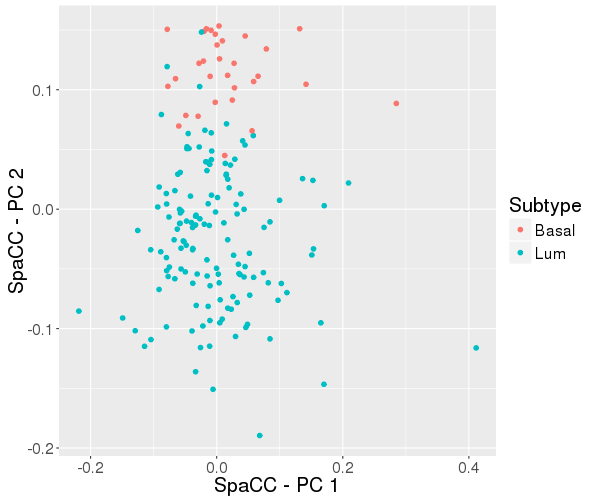}  &
\includegraphics[width=.3\linewidth]{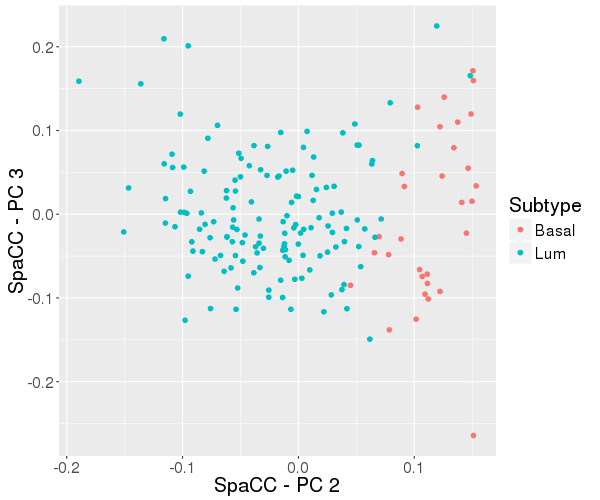} &
\multirow{2}{*}[.6in]{\includegraphics[height=.15\textheight,width=.35\linewidth]{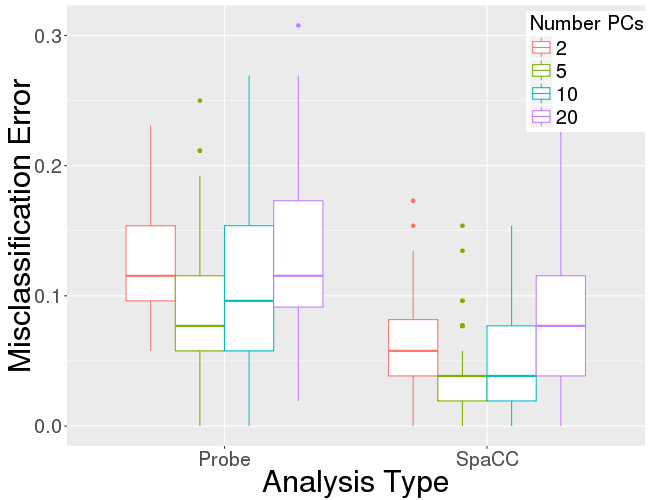}}\\
\includegraphics[width=.3\linewidth]{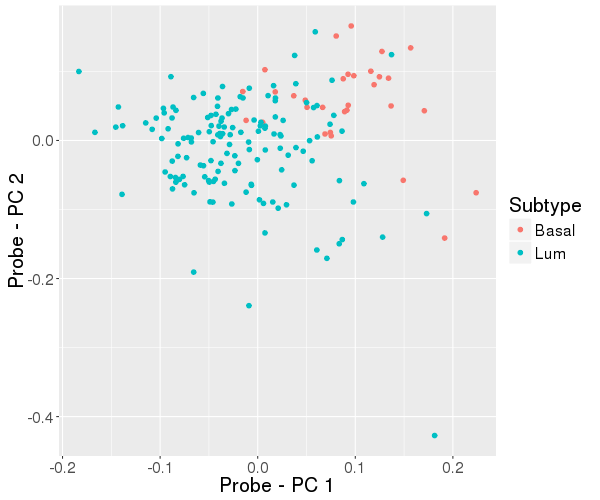}  &
\includegraphics[width=.3\linewidth]{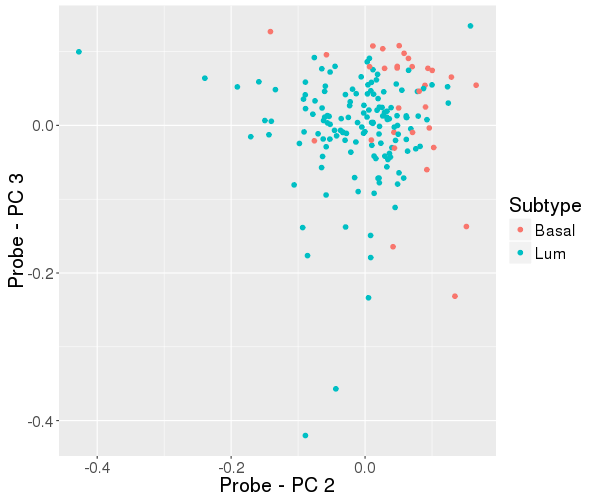}
\end{tabular}
\caption{ (Left) Principal Component scatterplots for SpaCC features
  (top) and raw probes (bottom), illustrating increased separation
  between Basal and Luminal subtypes for SpaCC features. (Right) Naive
  Bayes misclassification error on repeated test/train splits for both
  Probe and SpaCC features. SpaCC achieves lower error rates for all
  number of PC's.}
\label{figure:pcafig}
\end{figure}

Several have recently suggested that methylation levels can be used to
define cancer subtypes, and in breast cancer, methylation levels
have been used to characterize the well-known expression-based
subtypes \citep{van2010array}.  Here, we illustrate how reducing
methylation data to the SpaCC-derived reduced set of features offers
improvements in downstream multivariate analyses such as subtype
detection.  We continue to work with the TCGA breast cancer data for
chromosome 17 and consider classifying between the Basal and Luminal
(A and B) subtypes.  To fairly assess the performance of SpaCC's
region-based features relative to using the raw probe values, we
consider a simple classification scheme where we first reduce the data
using principal components and then fit a Naive Bayes classifier.
In Figure~\ref{figure:pcafig}, we repeatedly split the data into
training and test sets and report the misclassification errors on the
test sets for a range of principal components (PCs).  We also show PC
scatterplots for SpaCC-based and probe-based analysis, illustrating
the superior discriminatory power of SpaCC-based features.  Also
notice that SpaCC-based features outperform in terms of
misclassification error for all the number of PCs considered.  By
reducing the noisy probe-level data into more biologically meaningful
units, SpaCC is able to yield improved results for subtype detection.

\subsubsection{Inferring Epigenetic Networks}
\label{subsubsection:appmethnetwork}

Various genomic regions have been consistently shown to be hyper-
or hypomethylated together in sets of cancer patients ~\citep{esteller2001gene}.
One way of understanding relationships between methylated regions in
cancer is by inferring epigenetic networks.  Here, we show how SpaCC's
region-based features can lead to more meaningful and interpretable
epigenetic networks.  We continue to study the TCGA breast cancer
chromosome 17 data and use Gaussian Graphical Models to infer networks
where either the raw probes or the SpaCC-estimated regions serve as nodes.
Networks are estimated via the graphical lasso ~\citep{friedman2008sparse}
with a common (dimension dependent) regularization parameter $\lambda
= 8\sqrt{\frac{\log(p)}{n}}$ and stability selection
\citep{meinshausen2010stability} with a common threshold parameter of
$\tau = .95$; we utilize the BigQUIC software ~\citep{dhillonbig} to
estimate both the probe and region-based networks.

\begin{figure}[H]
  \centerline{
  \includegraphics[clip, trim=0.5cm 15cm 0.5cm 0.0cm, width=.7\linewidth]{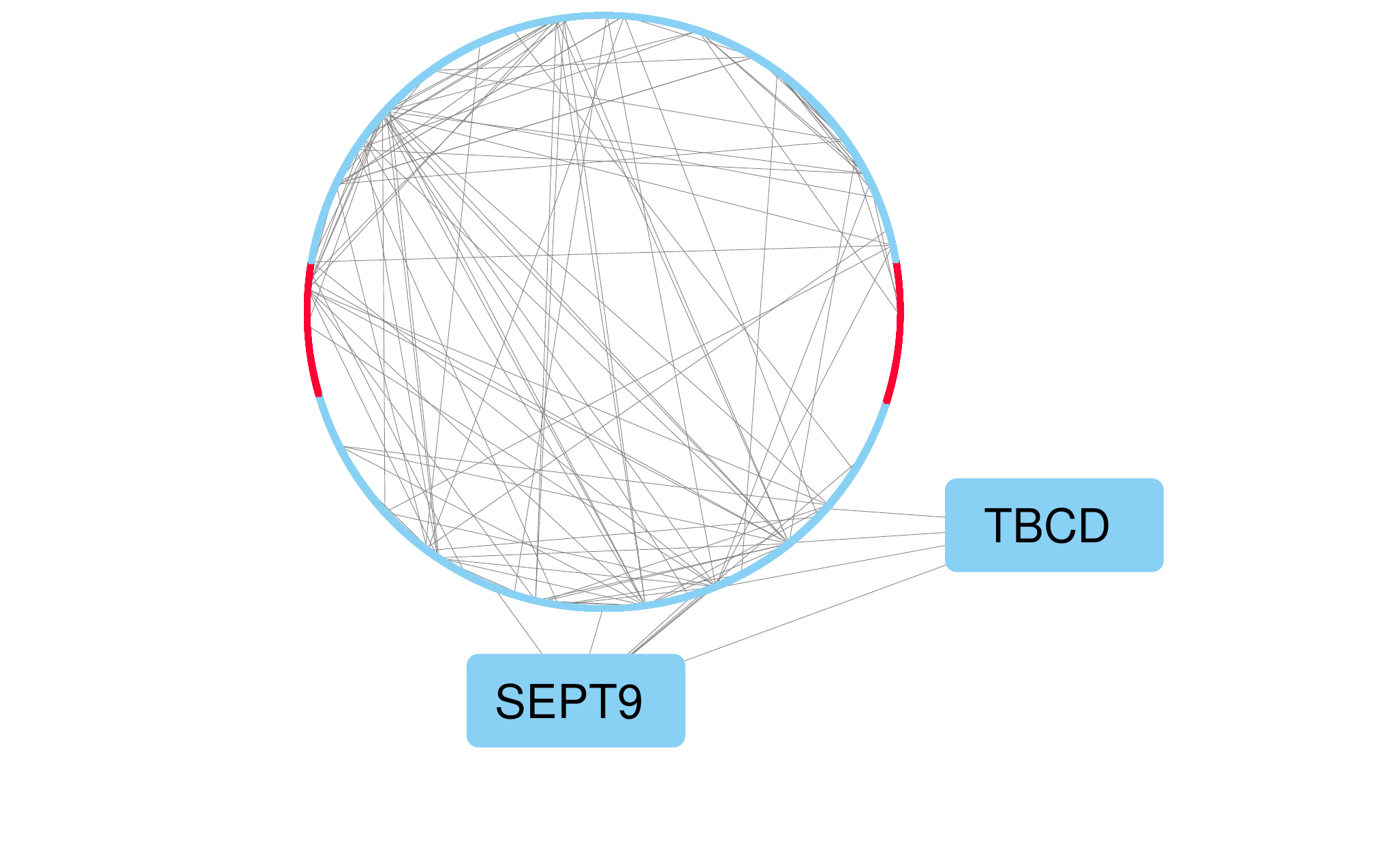}%
  \hspace{-1cm}
  \includegraphics[clip, trim=0.5cm 15cm 0.5cm 0.0cm, width=.7\linewidth]{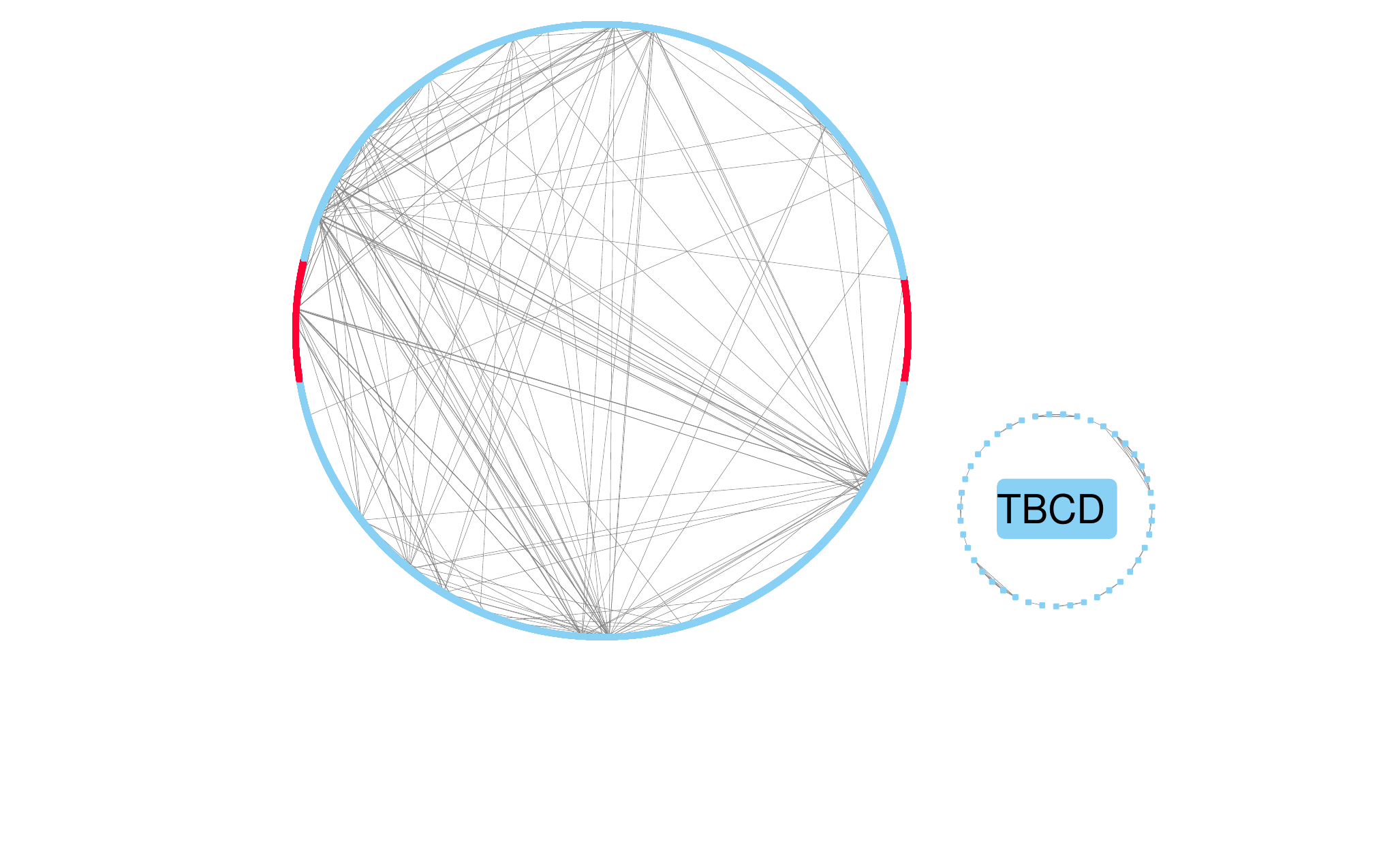}
  }
  \caption{Estimated methylation networks for Chromosome 17. The presence of uninformative short-range regional connections obtained with raw probe data (right) is illustrated via the TBCD gene. The region-based network (left) eliminates local connections allowing for longer range relationships (e.g. SEPT9 gene)}
  \label{fig:ggm}
\end{figure}

Results for the probe-based and region-based epigenetic networks are
shown in
Figure \ref{fig:ggm}.  Networks
inferred for probe features contain a much larger proportion of edges
connecting spatially adjacent probes; summaries of the genomic
distances between connected nodes are given in Table~\ref{tab:dist}.
This finding is not unexpected given
the high degree of spatial correlation inherent in methylation
levels.  Connections between local regions, however, are less
meaningful biologically.  SpaCC, on the other hand, reduces the data
to its relevant biological units and hence inferred networks have a
larger portion of long range connections that are more likely of
interest to scientists.  For example, the TBCD gene highlighted in
Figure~\ref{fig:ggm} plays a role in the cytokinesis stage of cell division
~\cite{fanarraga2010tbcd} and has been shown to be upregulated in
breast cancer ~\cite{folgueira2005gene}.   As can be seen in the
probe-based inferred network, probes in the TBCD region of chromosome 17
form only local connections.
In contrast, SpaCC aggregates many of the probes in the regions
surrounding the TBCD gene.
The resulting network estimate no longer contains TBCD
intra-gene connections, and instead forms longer range connections.
One such long range connection illustrated above is to the SEPT9 gene region
which has exhibited high expression levels in breast cancer cell lines
~\cite{montagna2003septin}.   While the connection between the epigenetics
of TBCD and SEPT9 has not been experimentally established, both genes
have been associated with breast cancer and thus represent the
precise type of potential connection which scientists seek to discover
using graphical models.  By aggregating probes into relevant
biological units, SpaCC-based graphical models are
able to discover potentially novel functional relationships between
epigenomic regions which would otherwise be masked.

\begin{table}[H]
\begin{center}
\hspace*{-.3in}\begin{tabular}{| l | l | l | l | l | l | l | l |}
\hline
Analysis Type & Mean & Median & 25\% & 75\% & 90\% & 95\% \\
\hline
\cellcolor{Tgray} Region & 6554.18 & 30.33 & .18 & 5252.73 &  31371.98 & 36738.76  \\
\cellcolor{Tgray} Probe & 2589.0428 & 0.075 & .019  & .208  & 2509.20 & 21967.56 \\
\hline
\end{tabular}
\caption{Summary of genomic distances between connected nodes for both SpaCC features and probe features, measured in kb. We note SpaCC favors longer range connections, as evidenced by larger genomic distance between neighbors.}
\label{tab:dist}
\end{center}
\end{table}

\subsection{Region-Based Epigenetic-Wide Association Studies}
\label{subsection:apprewas}

Finally, we study how SpaCC can improve systematic discovery of
epigentic markers associated with disease through epigenome-wide
association studies (EWAS).   Analogous to more widely used GWAS, EWAS
conducts univariate tests for association with an outcome at each
epigenetic marker (probe at a CpG site) and adjusts significance levels for
multiplicity.  As the epigenome can encode environmental or behavioral
characteristics of human subjects ~\citep{breitling2011tobacco}, EWAS
can help discover how factors other than genetics contribute to
disease.  For EWAS with DNA Methylation data, however, the number of
subjects is typically small relative to the $\approx 450,000$ CpG
sites measured by the latest Illumina platform, thus leading to a
possibly under-powered study.  To address this, we propose to first
reduce methylation data to genomic regions via SpaCC and then conduct
an association study on the regions; we term this a region-based
epigenome-wide association study (rEWAS).  As SpaCC
retains genomic regions that behave as functional units, we expect that
this will reduce the number of tests conducted, leading to an increase
in statistical power, while still being able to detect epigenetic
markers of disease.  Note that testing regions in rEWAS is similar in
spirit to testing groups of genetic markers based on linkage
disequilibrium in GWAS \citep{welter2014nhgri}.
Also note that some EWAS analyses have proposed to test regions, but
they have typically considered tests for global methylation levels
~\citep{hsiung2007global,ogino2008cohort} instead of localized genomic
regions as we propose with SpaCC.  In this section, we first evaluate
the efficacy of SpaCC-based rEWAS via a simulation study and then
present a rEWAS example to detect epigenetic markers of survival in
lung cancer.

\subsubsection{rEWAS Simulation Study}
\label{subsubsection:apprewassim}

Since rEWAS is a new type of association study, we first use simulated
methylation data to assess the performance of SpaCC-based rEWAS
compared to traditional EWAS analysis using the raw probes.  As in
Section~\ref{ref:methsim}, we base our simulated data on TCGA Breast
Cancer Level 3 methylation data for chromosome 17.  We obtain initial
regions, $\left\{ c_g \right\}$ via SpaCC and then simulate region
means for each subject as 
$m_{ig} \sim beta(2,2)$. Next, individual probes, $X_{ij}$ are
simulated as deviations about the region means via $X_{ij} \sim
beta(\alpha,\tau)$, where $\alpha,\tau$ are chosen to ensure
$E[X_{ij}] = m_{ig}$. Finally, we generate our response as a linear
function of a subset of the region means: $y_i = \beta_1 m_{ig_1} +
\dots \beta_d m_{ig_d} + \epsilon_i$, where $\epsilon \sim
N(0,\sigma^2)$, and $\beta_l \sim
N(\beta_{seed},\sqrt{\beta_{seed}})$. The difficulty of the simulation
is controlled by the signal-to-ratio (SNR) level, here the size of
$\beta_{seed}$ relative to $\sigma^2$.  Our simulation consists of
$n=94$ patients relative to $p=23,515$ probes.

We compare our SpaCC-based rEWAS approach to rEWAS using linkage
disequilibrium ~\citep{shoemaker2010allele} and the Fisher product method
~\citep{yu2015association} as well as EWAS on the raw probes.  
For all methods, we fit a univariate linear regression model at each
probe or region and corrected for multiplicity using
Benjamini-Hochberg's method to control the FDR \citep{benjamini1995controlling}.
The objective is to recover the regions or all the probes contained
within the regions that determine the 
outcome.  Notice then that there are two ways to report the true
positive rate (TPR) and false discovery proportion (FDP) for this
simulation study.  First, we can use the regions detected as
significant and compare the spatial extent of a detected region to
that of the corresponding true region to determine a true positive
rate; we call this the region point-of-view (RegionPOV) and the FDP is
defined analogously.  Second, we can compare the probes detected, or
the probes within regions detected as significant, to the probes that
lie within the true regions to determine the true positive rate; we
call this the probe point-of-view (ProbePOV). The Supplemental
Materials contain formal definitions of these metrics used to evaluate
our simulation study.  

\begin{figure}[H]
  \centerline{
  \includegraphics[width=1\linewidth]{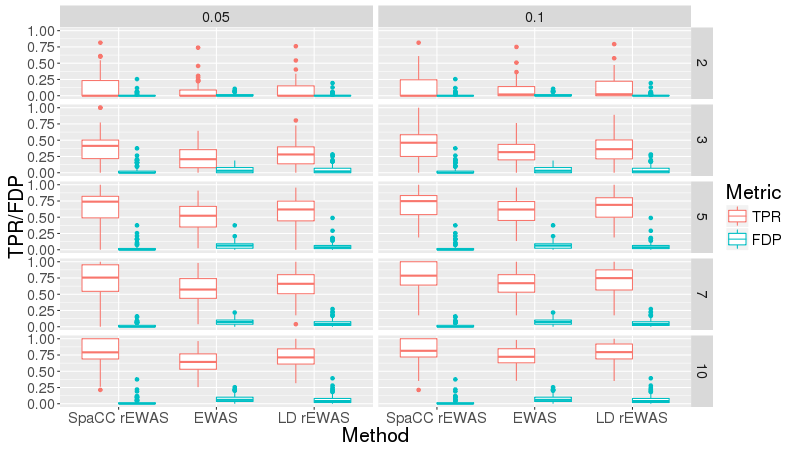}%
  }
  \centerline{
  \includegraphics[width=1\linewidth]{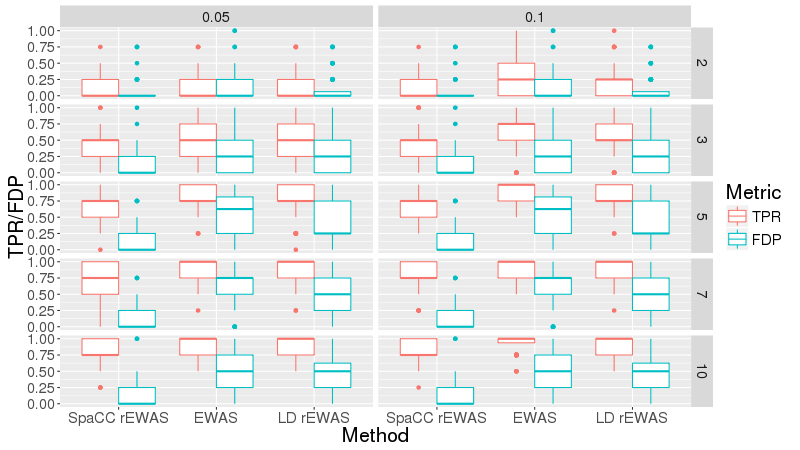}%
  }
  \caption{rEWAS Simulation Results: True Positive Rate (TPR) and
    False Discovery Proportion (FDP) for SpaCC-based rEWAS, LD-based rEWAS
    and probe-based EWAS over various SNR levels and FDR levels.
    Results are reported according to both probe point-of-view
    (ProbePOV; top) and Region point-of-view (RegionPOV; bottom)
    metrics.} 
  \label{fig:rewassim}
\end{figure}

In Figure ~\ref{fig:rewassim}, we report the results of our simulation
study at various SNR levels and FDR levels.  Compared to probe-based
EWAS and LD-based rEWAS, SpaCC-based rEWAS, achieves comparable FDP levels
while achieving higher TPR according to both probe and region-based
metrics.  Overall, this study confirms our intuition that first
reducing methylation data to regions via SpaCC, leads to increases in
statistical power for EWAS.

\subsubsection{Lung Cancer rEWAS Study}
\label{subsubsection:apprewaslung}

\begin{figure}[H]
  \centerline{
  \includegraphics[width=.8\linewidth]{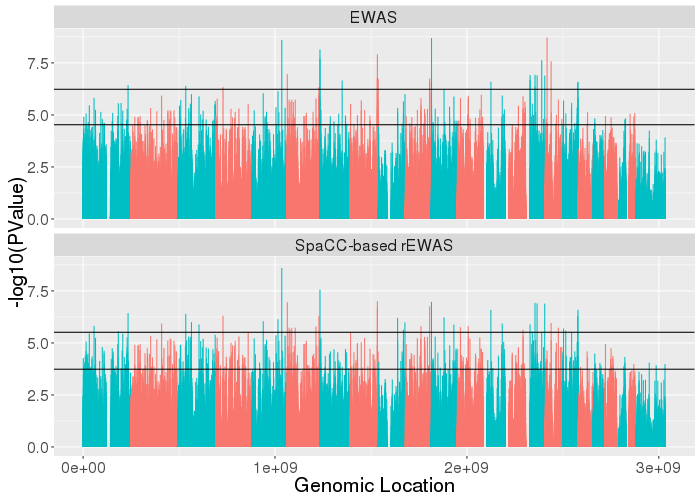}
  }
  \caption{Manhattan Plots with 1\% and 5\% FDR lines for EWAS (top) and SpaCC-based rEWAS (bottom) analysis for lung cancer methylation. We note the region-based analysis recovers a sizable portion of the probes deemed significant in the probe-based analysis. In addition, a large number of additional probes are recovered at an equivalent FDR-level.}
  \label{fig:manhattan}
\end{figure}

We now use our SpaCC-based rEWAS approach to discover epigenetic
markers associated with lung cancer survival.  We use the TCGA lung
cancer DNA methylation data which has $n =458$ patients and
$p=394,001$ probes across all
chromosomes~\citep{cancer2014comprehensive}.  A univariate Cox
proportional hazards model is used to test for associations with
survival at each probe (CpG site) for EWAS or at each SpaCC-region for
rEWAS.  The Benjamini-Hochberg procedure was used to control the FDR
at the 1\% and 5\% levels.  The p-values at each genomic
location for EWAS and SpaCC-based rEWAS are displayed in Manhattan plots in
Figure~\ref{fig:manhattan}; the alternating colors represent
chromosomes.   Horizontal lines are shown at the 1\% and 5\% FDR
levels; vertical lines denoting p-values that cross these
thresholds are statistically significant.  Notice that the Manhattan
plots for both EWAS and rEWAS retain a similar shape, indicating that
both methods found a common epigenome-wide signature for lung cancer
survival. On closer inspection, we see that SpaCC-based rEWAS yields a
larger number of discoveries at equivalent FDR-levels.  At the 1\% FDR
level, EWAS found 29 significant CpG sites, whereas SpaCC-based rEWAS
found 49 significant regions which contain a total of 77 CpG sites.
Similarly at the 5\% FDR level, EWAS had 287 discoveries while
SpaCC-based rEWAS found 556 regions which contain 861 CpG sites.
The overlap in significant discoveries are shown in Table
~\ref{tab:counts}.  Notice especially at the 1\% FDR level that our
SpaCC rEWAS method missed only 4 CpG sites declared significant by
EWAS, but in the converse, EWAS missed 54 discoveries made by rEWAS.  

\begin{table}[H]
\hspace*{.8in}
\begin{tabular}{| l | l | l | l |}
\hline
\multicolumn{2}{|c|}{} & \multicolumn{2}{c|}{Region}\\
\cline{3-4}
\multicolumn{2}{|c|}{ } & $+$ & $-$\\
\hline
\multirow{2}{*}{Probe} & $+$ & 247 (216) & 40 (39) \\
\cline{2-4}
& $-$ & \cellcolor{Tgray} 614 (340) &  \\
\hline
\end{tabular}
\quad
\quad
\quad
\quad
\begin{tabular}{| l | l | l | l |}
\hline
\multicolumn{2}{|c|}{} & \multicolumn{2}{c|}{Region}\\
\cline{3-4}
\multicolumn{2}{|c|}{ } & $+$ & $-$\\
\hline
\multirow{2}{*}{Probe} & $+$ & 25 (21) & 4 (4) \\
\cline{2-4}
& $-$ &\cellcolor{Tgray} 52 (28) &  \\
\hline
\end{tabular}
\caption{(Left) Probe (region) significance overlap between region and probe based analysis at 5\% FDR level. Total Number of .05-level probes is 287. Total Number of .05-level regions is 556. Total Number of probes in .05-level regions is 861.(Right) Similarly at 1\% FDR level. Total number of .01-level probes is 29 . Total number of .01-level regions is 49. Total Number of probes in .01-regions is 77.
}
\label{tab:counts}
\end{table}

Our analysis reveals several important epigenetic markers which are
largely consistent with the cancer literature.  In the top portion of
the Table \ref{tab:disc}, we list the most significant discoveries
from rEWAS which were also discovered by EWAS.  We also report the
epigenetic marker location, its gene target or the nearest gene, and a
brief description of its role in the cancer literature.  Note that
further details on the literature are given in the Supplemental
Materials.  In the bottom portion of Table~\ref{tab:disc}, we
highlight the most significant discoveries found exclusively by
rEWAS; these are known markers for lung cancer.  Overall, our analysis
reveals that reducing methylation data to biologically meaningful
genomic regions via SpaCC before conducting EWAS studies, leads to
major increases in statistical power for discovering
epigenetic markers.

\begin{table}[H]
\begin{center}
\hspace*{-.3in}
\scalebox{0.8}{
\begin{tabular}{| l | l | l | l |}
\hline
Gene & p-value & Chrm.(Loc.) & Description \\
\hline
\cellcolor{Tgray} LARP1 & 2.5e-9 & 5 (154.09-154.197)  & Regulator of mTOR, prognostic marker \\
\cellcolor{Tgray} ZFAND2A & 2.7e-8 & 7 (1.198 - 1.199) & Target for lung cancer therapy \\
\cellcolor{Tgray} TRAPPC9 & 9.8e-8 & 8 (140.74 - 141.46) & High expression in cancer cell lines  \\
\cellcolor{Tgray} PKP3 & 1.0e-7 & 11 (.39 - .40) & Oncogene,prognostic marker \\
\cellcolor{Tgray} GMDS & 1.1e-7 & 6 (1.62 - 2.24) & Relation to NK escape \\
\cellcolor{Tgray} FBN1 & 1.1e-7 & 15 (48.70 - 48.93) &  Hypermethylation in colorectal cancer \\
\cellcolor{Tgray} MYO1E & 1.2e-7 & 15 (59.42 - 59.66)  & Inhibition may prevent metastasis \\
\cellcolor{Tgray} IGF1R & 1.2e-7 & 15 (99.19 - 99.50)  & Silencing enhances sensitivity to DNA-damage \\
\cellcolor{Tgray} FAM53B & 1.7e-7 & 10 (126.30 - 126.43) & Role in cell proliferation \\
\cellcolor{Tgray} ANAPC11 & 2.2e-7 & 17 (79.84 - 79.85) & Role in lung development. \\
\hline
\hline
\hline
 \cellcolor{Tgray} {\color{red}CCDC12} & 7.6e-6 & 3 (46.96 - 47.02) & Contained in 3p21.3 tumor suppressor region \\
 \cellcolor{Tgray} {\color{red}WWOX} & 1.6e-5 & 16 (78.13 -79.24) & Biomarker for lung cancer \\
 \cellcolor{Tgray} {\color{red}ARL14} & 1.8e-5 & 5 (160.394 - 160.396)  & Homologue to tumor suppressor gene ARLTS1 \\
\hline
\end{tabular}}
\caption{Significant Discoveries found by SpaCC-based rEWAS. The top
  portion describes the top ten most significant discoveries; the
  bottom portion (in red) describes a subset of discoveries detected
  exclusively by rEWAS. } 
\label{tab:disc}
\end{center}
\end{table}

\section{Discussion}

Building on the success of fusion-based penalties and convex
clustering methods, 
we have introduced a clustering technique for spatially correlated
data called Spatial Convex Clustering, or SpaCC.   
Through both simulations and real data examples we have shown SpaCC to
be a successful tool for reducing DNA methylation data to its
functional genomic regions.  The resulting regions can then be used to
discover epigenetic markers or as features for down stream
multivariate statistical analyses. 
While this paper has focused on the analysis of methylation data, the
SpaCC framework is not limited to this data type, nor genomic data in
particular.  Extensions of SpaCC may be appropriate for clustering
spatially registered read counts from RNA-sequencing data to discover
new isoforms and alternate splicing.  Beyond genomics,
similar fusion-based approaches may also be applied other biological
data sources with known local spatial structure such as brain
imaging.  Taken together, these future extensions can allow scientists
to perform a single integrative analysis to discover meaningful regions
across differing biological modalities.  An {\tt R} package {\tt
  SpaCCr} implementing our method is available from CRAN ~\citep{spaccr}.

\section*{Acknowledgements}

J.N. was supported in this work by the NIH Training Award in
Biostatistics for Cancer Research Award 5T32CA09652010.
J.N. and G.I.A. were
supported by NFS DMS-1264058 and  NFS DMS-1554821. We also thank
Zhandong Liu and 
Ying-Wooi Wan at Baylor College of Medicine for thoughtful discussions
related to this work.

\bibliography{SpaCCPaper}

\begin{thebibliography}{48}
\providecommand{\natexlab}[1]{#1}
\providecommand{\url}[1]{\texttt{#1}}
\expandafter\ifx\csname urlstyle\endcsname\relax
  \providecommand{\doi}[1]{doi: #1}\else
  \providecommand{\doi}{doi: \begingroup \urlstyle{rm}\Url}\fi

\bibitem[Benjamini and Hochberg(1995)]{benjamini1995controlling}
Yoav Benjamini and Yosef Hochberg.
\newblock Controlling the false discovery rate: a practical and powerful
  approach to multiple testing.
\newblock \emph{Journal of the royal statistical society. Series B
  (Methodological)}, pages 289--300, 1995.

\bibitem[Bibikova et~al.(2009)Bibikova, Le, Barnes, Saedinia-Melnyk, Zhou,
  Shen, and Gunderson]{bibikova2009genome}
Marina Bibikova, Jennie Le, Bret Barnes, Shadi Saedinia-Melnyk, Lixin Zhou,
  Richard Shen, and Kevin~L Gunderson.
\newblock Genome-wide dna methylation profiling using infinium{\textregistered}
  assay.
\newblock 2009.

\bibitem[Bleakley and Vert(2011)]{bleakley2011group}
Kevin Bleakley and Jean-Philippe Vert.
\newblock The group fused lasso for multiple change-point detection.
\newblock \emph{arXiv preprint arXiv:1106.4199}, 2011.

\bibitem[Breitling et~al.(2011)Breitling, Yang, Korn, Burwinkel, and
  Brenner]{breitling2011tobacco}
Lutz~P Breitling, Rongxi Yang, Bernhard Korn, Barbara Burwinkel, and Hermann
  Brenner.
\newblock Tobacco-smoking-related differential dna methylation: 27k discovery
  and replication.
\newblock \emph{The American Journal of Human Genetics}, 88\penalty0
  (4):\penalty0 450--457, 2011.

\bibitem[Chi and Lange()]{ChiLan2014a}
Eric~C. Chi and Kenneth Lange.
\newblock Splitting methods for convex clustering.
\newblock \emph{Journal of Computational and Graphical Statistics}.
\newblock \doi{10.1080/10618600.2014.948181}.
\newblock URL \url{http://dx.doi.org/10.1080/10618600.2014.948181}.
\newblock In press.

\bibitem[Chi et~al.(2016)Chi, Allen, and Baraniuk]{chi2016convex}
Eric~C Chi, Genevera~I Allen, and Richard~G Baraniuk.
\newblock Convex biclustering.
\newblock \emph{Biometrics}, 2016.

\bibitem[Das and Singal(2004)]{das2004dna}
Partha~M Das and Rakesh Singal.
\newblock Dna methylation and cancer.
\newblock \emph{Journal of clinical oncology}, 22\penalty0 (22):\penalty0
  4632--4642, 2004.

\bibitem[Dhillon()]{dhillonbig}
Inderjit~S Dhillon.
\newblock Big \& quic: Sparse inverse covariance estimation for a million
  variables.

\bibitem[Eckhardt et~al.(2006)Eckhardt, Lewin, Cortese, Rakyan, Attwood,
  Burger, Burton, Cox, Davies, Down, et~al.]{eckhardt2006dna}
Florian Eckhardt, Joern Lewin, Rene Cortese, Vardhman~K Rakyan, John Attwood,
  Matthias Burger, John Burton, Tony~V Cox, Rob Davies, Thomas~A Down, et~al.
\newblock Dna methylation profiling of human chromosomes 6, 20 and 22.
\newblock \emph{Nature genetics}, 38\penalty0 (12):\penalty0 1378--1385, 2006.

\bibitem[Esteller et~al.(2001)Esteller, Corn, Baylin, and
  Herman]{esteller2001gene}
Manel Esteller, Paul~G Corn, Stephen~B Baylin, and James~G Herman.
\newblock A gene hypermethylation profile of human cancer.
\newblock \emph{Cancer research}, 61\penalty0 (8):\penalty0 3225--3229, 2001.

\bibitem[Fanarraga et~al.(2010)Fanarraga, Bellido, Ja{\'e}n, Villegas, and
  Zabala]{fanarraga2010tbcd}
M{\'o}nica~L{\'o}pez Fanarraga, Javier Bellido, Cristina Ja{\'e}n, Juan~Carlos
  Villegas, and Juan~Carlos Zabala.
\newblock Tbcd links centriologenesis, spindle microtubule dynamics, and
  midbody abscission in human cells.
\newblock \emph{PloS one}, 5\penalty0 (1):\penalty0 e8846, 2010.

\bibitem[Folgueira et~al.(2005)Folgueira, Carraro, Brentani,
  da~Costa~Patr{\~a}o, Barbosa, Netto, Caldeira, Katayama, Soares, Oliveira,
  et~al.]{folgueira2005gene}
Maria Aparecida Azevedo~Koike Folgueira, Dirce~Maria Carraro, Helena Brentani,
  Diogo~Ferreira da~Costa~Patr{\~a}o, Edson~Mantovani Barbosa,
  M{\'a}rio~Mour{\~a}o Netto, Jos{\'e} Roberto~F{\'\i}garo Caldeira, Maria
  Lucia~Hirata Katayama, Fernando~Augusto Soares, C{\'e}lia~Tosello Oliveira,
  et~al.
\newblock Gene expression profile associated with response to doxorubicin-based
  therapy in breast cancer.
\newblock \emph{Clinical Cancer Research}, 11\penalty0 (20):\penalty0
  7434--7443, 2005.

\bibitem[Friedman et~al.(2008)Friedman, Hastie, and
  Tibshirani]{friedman2008sparse}
Jerome Friedman, Trevor Hastie, and Robert Tibshirani.
\newblock Sparse inverse covariance estimation with the graphical lasso.
\newblock \emph{Biostatistics}, 9\penalty0 (3):\penalty0 432--441, 2008.

\bibitem[Hansen et~al.(2012)Hansen, Langmead, Irizarry,
  et~al.]{hansen2012bsmooth}
Kasper~D Hansen, Benjamin Langmead, Rafael~A Irizarry, et~al.
\newblock Bsmooth: from whole genome bisulfite sequencing reads to
  differentially methylated regions.
\newblock \emph{Genome Biol}, 13\penalty0 (10):\penalty0 R83, 2012.

\bibitem[Hocking et~al.(2011)Hocking, Vert, Joulin, and
  Bach]{hocking2011clusterpath}
Toby Hocking, Jean-Philippe Vert, Armand Joulin, and Francis~R Bach.
\newblock Clusterpath: an algorithm for clustering using convex fusion
  penalties.
\newblock In \emph{Proceedings of the 28th International Conference on Machine
  Learning (ICML-11)}, pages 745--752, 2011.

\bibitem[Hsiung et~al.(2007)Hsiung, Marsit, Houseman, Eddy, Furniss, McClean,
  and Kelsey]{hsiung2007global}
Debra~Ting Hsiung, Carmen~J Marsit, E~Andres Houseman, Karen Eddy, C~Sloane
  Furniss, Michael~D McClean, and Karl~T Kelsey.
\newblock Global dna methylation level in whole blood as a biomarker in head
  and neck squamous cell carcinoma.
\newblock \emph{Cancer Epidemiology Biomarkers \& Prevention}, 16\penalty0
  (1):\penalty0 108--114, 2007.

\bibitem[Hunter and Lange(2004)]{hunter2004tutorial}
David~R Hunter and Kenneth Lange.
\newblock A tutorial on mm algorithms.
\newblock \emph{The American Statistician}, 58\penalty0 (1):\penalty0 30--37,
  2004.

\bibitem[Jaffe et~al.(2012)Jaffe, Murakami, Lee, Leek, Fallin, Feinberg, and
  Irizarry]{jaffe2012bump}
Andrew~E Jaffe, Peter Murakami, Hwajin Lee, Jeffrey~T Leek, M~Daniele Fallin,
  Andrew~P Feinberg, and Rafael~A Irizarry.
\newblock Bump hunting to identify differentially methylated regions in
  epigenetic epidemiology studies.
\newblock \emph{International journal of epidemiology}, 41\penalty0
  (1):\penalty0 200--209, 2012.

\bibitem[King et~al.(2003)King, Marks, Mandell, et~al.]{king2003breast}
Mary-Claire King, Joan~H Marks, Jessica~B Mandell, et~al.
\newblock Breast and ovarian cancer risks due to inherited mutations in brca1
  and brca2.
\newblock \emph{Science}, 302\penalty0 (5645):\penalty0 643--646, 2003.

\bibitem[Meil{\u{a}}(2007)]{meilua2007comparing}
Marina Meil{\u{a}}.
\newblock Comparing clusterings�an information based distance.
\newblock \emph{Journal of multivariate analysis}, 98\penalty0 (5):\penalty0
  873--895, 2007.

\bibitem[Meinshausen and B{\"u}hlmann(2010)]{meinshausen2010stability}
Nicolai Meinshausen and Peter B{\"u}hlmann.
\newblock Stability selection.
\newblock \emph{Journal of the Royal Statistical Society: Series B (Statistical
  Methodology)}, 72\penalty0 (4):\penalty0 417--473, 2010.

\bibitem[Meinshausen and Yu(2009)]{meinshausen2009lasso}
Nicolai Meinshausen and Bin Yu.
\newblock Lasso-type recovery of sparse representations for high-dimensional
  data.
\newblock \emph{The Annals of Statistics}, pages 246--270, 2009.

\bibitem[Montagna et~al.(2003)Montagna, Lyu, Hunter, Lukes, Lowther, Reppert,
  Hissong, Weaver, and Ried]{montagna2003septin}
Cristina Montagna, Myung-Soo Lyu, Kent Hunter, Luanne Lukes, William Lowther,
  Tricia Reppert, Bruce Hissong, Zo{\"e} Weaver, and Thomas Ried.
\newblock The septin 9 (msf) gene is amplified and overexpressed in mouse
  mammary gland adenocarcinomas and human breast cancer cell lines.
\newblock \emph{Cancer research}, 63\penalty0 (9):\penalty0 2179--2187, 2003.

\bibitem[Nagorski()]{spaccr}
John Nagorski.
\newblock \emph{SpaCCr: A package for genomic region detection via spatial
  convex clustering.}
\newblock R package version 0.1.0.

\bibitem[Network et~al.(2012)]{cancer2012comprehensive}
Cancer Genome~Atlas Network et~al.
\newblock Comprehensive molecular portraits of human breast tumours.
\newblock \emph{Nature}, 490\penalty0 (7418):\penalty0 61--70, 2012.

\bibitem[Network et~al.(2011)]{cancer2011integrated}
Cancer Genome Atlas~Research Network et~al.
\newblock Integrated genomic analyses of ovarian carcinoma.
\newblock \emph{Nature}, 474\penalty0 (7353):\penalty0 609--615, 2011.

\bibitem[Network et~al.(2014)]{cancer2014comprehensive}
Cancer Genome Atlas~Research Network et~al.
\newblock Comprehensive molecular profiling of lung adenocarcinoma.
\newblock \emph{Nature}, 511\penalty0 (7511):\penalty0 543--550, 2014.

\bibitem[Nowak et~al.(2011)Nowak, Hastie, Pollack, and
  Tibshirani]{nowak2011fused}
Gen Nowak, Trevor Hastie, Jonathan~R Pollack, and Robert Tibshirani.
\newblock A fused lasso latent feature model for analyzing multi-sample acgh
  data.
\newblock \emph{Biostatistics}, page kxr012, 2011.

\bibitem[O'Brien et~al.(2008)O'Brien, Cavet, Pandita, Hu, Haydu, Mohan, Toy,
  Rivers, Modrusan, Amler, et~al.]{o2008functional}
Carol O'Brien, Guy Cavet, Ajay Pandita, Xiaolan Hu, Lauren Haydu, Sankar Mohan,
  Karen Toy, Celina~Sanchez Rivers, Zora Modrusan, Lukas~C Amler, et~al.
\newblock Functional genomics identifies abcc3 as a mediator of taxane
  resistance in her2-amplified breast cancer.
\newblock \emph{Cancer research}, 68\penalty0 (13):\penalty0 5380--5389, 2008.

\bibitem[Ogino et~al.(2008)Ogino, Nosho, Kirkner, Kawasaki, Chan, Schernhammer,
  Giovannucci, and Fuchs]{ogino2008cohort}
Shuji Ogino, Katsuhiko Nosho, Gregory~J Kirkner, Takako Kawasaki, Andrew~T
  Chan, Eva~S Schernhammer, Edward~L Giovannucci, and Charles~S Fuchs.
\newblock A cohort study of tumoral line-1 hypomethylation and prognosis in
  colon cancer.
\newblock \emph{Journal of the National Cancer Institute}, 100\penalty0
  (23):\penalty0 1734--1738, 2008.

\bibitem[Olshen et~al.(2004)Olshen, Venkatraman, Lucito, and
  Wigler]{olshen2004circular}
Adam~B Olshen, ES~Venkatraman, Robert Lucito, and Michael Wigler.
\newblock Circular binary segmentation for the analysis of array-based dna copy
  number data.
\newblock \emph{Biostatistics}, 5\penalty0 (4):\penalty0 557--572, 2004.

\bibitem[Redon et~al.(2006)Redon, Ishikawa, Fitch, Feuk, Perry, Andrews,
  Fiegler, Shapero, Carson, Chen, et~al.]{redon2006global}
Richard Redon, Shumpei Ishikawa, Karen~R Fitch, Lars Feuk, George~H Perry,
  T~Daniel Andrews, Heike Fiegler, Michael~H Shapero, Andrew~R Carson, Wenwei
  Chen, et~al.
\newblock Global variation in copy number in the human genome.
\newblock \emph{nature}, 444\penalty0 (7118):\penalty0 444--454, 2006.

\bibitem[Seshan and Olshen()]{dnacopy}
Venkatraman~E. Seshan and Adam Olshen.
\newblock \emph{DNAcopy: DNA copy number data analysis}.
\newblock R package version 1.40.0.

\bibitem[Shlien and Malkin(2009)]{shlien2009copy}
Adam Shlien and David Malkin.
\newblock Copy number variations and cancer.
\newblock \emph{Genome medicine}, 1\penalty0 (6):\penalty0 1, 2009.

\bibitem[Shoemaker et~al.(2010)Shoemaker, Deng, Wang, and
  Zhang]{shoemaker2010allele}
Robert Shoemaker, Jie Deng, Wei Wang, and Kun Zhang.
\newblock Allele-specific methylation is prevalent and is contributed by
  cpg-snps in the human genome.
\newblock \emph{Genome research}, 20\penalty0 (7):\penalty0 883--889, 2010.

\bibitem[Taylor et~al.(2008)Taylor, Barretina, Socci, DeCarolis, Ladanyi,
  Meyerson, Singer, and Sander]{taylor2008functional}
Barry~S Taylor, Jordi Barretina, Nicholas~D Socci, Penelope DeCarolis, Marc
  Ladanyi, Matthew Meyerson, Samuel Singer, and Chris Sander.
\newblock Functional copy-number alterations in cancer.
\newblock \emph{PloS one}, 3\penalty0 (9):\penalty0 e3179, 2008.

\bibitem[Tibshirani et~al.(2005)Tibshirani, Saunders, Rosset, Zhu, and
  Knight]{tibshirani2005sparsity}
Robert Tibshirani, Michael Saunders, Saharon Rosset, Ji~Zhu, and Keith Knight.
\newblock Sparsity and smoothness via the fused lasso.
\newblock \emph{Journal of the Royal Statistical Society: Series B (Statistical
  Methodology)}, 67\penalty0 (1):\penalty0 91--108, 2005.

\bibitem[Troyanskaya et~al.(2001)Troyanskaya, Cantor, Sherlock, Brown, Hastie,
  Tibshirani, Botstein, and Altman]{troyanskaya2001missing}
Olga Troyanskaya, Michael Cantor, Gavin Sherlock, Pat Brown, Trevor Hastie,
  Robert Tibshirani, David Botstein, and Russ~B Altman.
\newblock Missing value estimation methods for dna microarrays.
\newblock \emph{Bioinformatics}, 17\penalty0 (6):\penalty0 520--525, 2001.

\bibitem[Tseng(1991)]{tseng1991applications}
Paul Tseng.
\newblock Applications of a splitting algorithm to decomposition in convex
  programming and variational inequalities.
\newblock \emph{SIAM Journal on Control and Optimization}, 29\penalty0
  (1):\penalty0 119--138, 1991.

\bibitem[van~de Geer et~al.(2011)van~de Geer, B{\"u}hlmann, Zhou,
  et~al.]{van2011adaptive}
Sara van~de Geer, Peter B{\"u}hlmann, Shuheng Zhou, et~al.
\newblock The adaptive and the thresholded lasso for potentially misspecified
  models (and a lower bound for the lasso).
\newblock \emph{Electronic Journal of Statistics}, 5:\penalty0 688--749, 2011.

\bibitem[Van~der Auwera et~al.(2010)Van~der Auwera, Yu, Suo, Van~Neste,
  Van~Dam, Van~Marck, Pauwels, Vermeulen, Dirix, and Van~Laere]{van2010array}
Ilse Van~der Auwera, Wayne Yu, Liping Suo, Leander Van~Neste, Peter Van~Dam,
  Eric~A Van~Marck, Patrick Pauwels, Peter~B Vermeulen, Luc~Y Dirix, and
  Steven~J Van~Laere.
\newblock Array-based dna methylation profiling for breast cancer subtype
  discrimination.
\newblock \emph{PLoS One}, 5\penalty0 (9):\penalty0 e12616, 2010.

\bibitem[Venkatraman and Olshen(2007)]{venkatraman2007faster}
ES~Venkatraman and Adam~B Olshen.
\newblock A faster circular binary segmentation algorithm for the analysis of
  array cgh data.
\newblock \emph{Bioinformatics}, 23\penalty0 (6):\penalty0 657--663, 2007.

\bibitem[Wagner and Wagner(2007)]{wagner2007comparing}
Silke Wagner and Dorothea Wagner.
\newblock \emph{Comparing clusterings: an overview}.
\newblock Universit{\"a}t Karlsruhe, Fakult{\"a}t f{\"u}r Informatik Karlsruhe,
  2007.

\bibitem[Wang et~al.(2005)Wang, Kim, Pollack, Narasimhan, and
  Tibshirani]{wang2005method}
Pei Wang, Young Kim, Jonathan Pollack, Balasubramanian Narasimhan, and Robert
  Tibshirani.
\newblock A method for calling gains and losses in array cgh data.
\newblock \emph{Biostatistics}, 6\penalty0 (1):\penalty0 45--58, 2005.

\bibitem[Welter et~al.(2014)Welter, MacArthur, Morales, Burdett, Hall, Junkins,
  Klemm, Flicek, Manolio, Hindorff, et~al.]{welter2014nhgri}
Danielle Welter, Jacqueline MacArthur, Joannella Morales, Tony Burdett, Peggy
  Hall, Heather Junkins, Alan Klemm, Paul Flicek, Teri Manolio, Lucia Hindorff,
  et~al.
\newblock The nhgri gwas catalog, a curated resource of snp-trait associations.
\newblock \emph{Nucleic acids research}, 42\penalty0 (D1):\penalty0
  D1001--D1006, 2014.

\bibitem[Wold(1978)]{wold1978cross}
Svante Wold.
\newblock Cross-validatory estimation of the number of components in factor and
  principal components models.
\newblock \emph{Technometrics}, 20\penalty0 (4):\penalty0 397--405, 1978.

\bibitem[Yu et~al.(2015)Yu, Chibnik, Srivastava, Pochet, Yang, Xu, Kozubek,
  Obholzer, Leurgans, Schneider, et~al.]{yu2015association}
Lei Yu, Lori~B Chibnik, Gyan~P Srivastava, Nathalie Pochet, Jingyun Yang, Jishu
  Xu, James Kozubek, Nikolaus Obholzer, Sue~E Leurgans, Julie~A Schneider,
  et~al.
\newblock Association of brain dna methylation in sorl1, abca7, hla-drb5,
  slc24a4, and bin1 with pathological diagnosis of alzheimer disease.
\newblock \emph{JAMA neurology}, 72\penalty0 (1):\penalty0 15--24, 2015.

\bibitem[Zhang()]{cntools}
Jianhua Zhang.
\newblock \emph{CNTools: Convert segment data into a region by sample matrix to
  allow for other high level computational analyses.}
\newblock R package version 1.22.0.

\end{thebibliography}
\end{document}